\documentclass[aps,prd,eqsecnum,11pt,showpacs,preprintnumbers,nofootinbib,superscriptaddress]{revtex4-1}
\usepackage{geometry}                
\usepackage[pdftex]{graphicx}
\usepackage{float}
\usepackage{amssymb,amsmath,amsopn,bm,dsfont}
\DeclareMathAlphabet{\matheul}{U}{eus}{m}{n}
\usepackage{setspace,enumitem}

\allowdisplaybreaks

\usepackage[colorlinks,pdfstartview=FitH]{hyperref}
\hypersetup{linkcolor=blue,citecolor=blue,filecolor=black,urlcolor=blue}

\newcommand{\gobble}[1]{}

\newcommand{\EE}{{\matheul E}}

\newcommand{\MM}{{\matheul M}}

\newcommand{\RH}{R_{_H}}
\newcommand{\RS}{R_{_S}}
\newcommand{\MKext}{M_K^{\mkern1mu\rm ext}}
\newcommand{\MKint}{M_K^{\mkern1mu\rm int}}
\newcommand{\MKsurface}{M_K^{\mkern1mu\rm surface}}
\newcommand{\MKorigin}{M_K^{\mkern1mu\rm origin}}
\newcommand{\MKvolume}{M_K^{\mkern1mu\rm volume}}
\newcommand{\JKext}{J_K^{\mkern1mu\rm ext}}
\newcommand{\JKint}{J_K^{\mkern1mu\rm int}}
\newcommand{\JKsurface}{J_K^{\mkern1mu\rm surface}}
\newcommand{\JKorigin}{J_K^{\mkern1mu\rm origin}}
\newcommand{\JKvolume}{J_K^{\mkern1mu\rm volume}}

\renewcommand\a{\alpha}
\renewcommand\b{\beta}
\renewcommand\d{\delta}
\renewcommand\k{\kappa}

\renewcommand\r{\rho}

\renewcommand\t{\tau}

\renewcommand\j{\psi}
\renewcommand\th{\theta}

\newcommand\e{\epsilon}

\newcommand\m{\mu}
\newcommand\n{\nu}

\newcommand\f{\phi}
\newcommand\w{\omega}

\renewcommand\S{\Sigma}

\newcommand\D{\Delta}

\newcommand\W{\Omega}

\newcommand{\cJ}{{\cal J}}

\newcommand{\cO}{{\cal O}}
\newcommand{\cS}{{\cal S}}

\newcommand{\pa}{\partial}

\newcommand{\nn}{\nonumber \\}
\newcommand{\na}{\nabla}
\newcommand{\sdfrac}[2]{\mbox{\small$\displaystyle\frac{#1}{#2}$}}

\newcommand{\bn}{{\boldsymbol n}}
\newcommand{\bN}{{\boldsymbol N}}

\def\nbox#1#2{\vcenter{\hrule \hbox{\vrule height#2in
\kern#1in \vrule} \hrule}}
\def\sq{\,\raise.5pt\hbox{$\nbox{.10}{.10}$}\,}
\def\sqb{\,\raise.5pt\hbox{$\overline{\nbox{.09}{.09}}$}\,}

\allowdisplaybreaks

\usepackage{xcolor}
\definecolor{markcolor}{rgb}{1.0, 0.13, 0.32}
\definecolor{mygray}{rgb}{0.31, 0.31, 0.61}

\begin{document}

\preprint{LA-UR-21-22546}

\title{Slowly rotating gravastars}
\author{Philip Beltracchi}
\email{philipbeltracchi@gmail.com}
\affiliation{Department of Physics and Astronomy, University of Utah\\Salt Lake City, Utah 84112, USA}

\author{Paolo Gondolo}
\email{paolo.gondolo@utah.edu}
\affiliation{Department of Physics and Astronomy, University of Utah\\Salt Lake City, Utah 84112, USA}
\affiliation{Department of Physics, Tokyo Institute of Technology, 2-12-1 Ookayama, Meguro-ku, Tokyo 152-8551, Japan}
\affiliation{Kavli Institute for the Physics and Mathematics of the Universe, The University of Tokyo, Kashiwa, Chiba 277-8583, Japan}

\author{Emil Mottola}
\email{mottola.emil@gmail.com, emottola@unm.edu}
\affiliation{Theoretical Division, T-2, MS B283, \\Los Alamos National Laboratory\\ Los Alamos, New Mexico 87545, USA}
\affiliation{Department of Physics and Astronomy, University of New Mexico\\
Albuquerque, New Mexico 87131, USA}

\begin{abstract}
\noindent We solve Einstein’s equations for slowly rotating gravitational condensate stars (gravastars) up to second order in the rotation by expanding about the spherically symmetric gravastar with de Sitter interior and Schwarzschild exterior matched at their common horizon.  Requiring that the perturbations are finite 
on the null surface reduces the exterior geometry to that of a Kerr black hole, implying that a slowly rotating gravastar cannot be distinguished from a Kerr black 
hole by any measurement or observation restricted to the macroscopic spacetime exterior to the horizon.  We determine the interior solution, the surface stress tensor, and the Komar mass and angular momentum localized on the slowly rotating horizon surface. 
With the interior equation of state fixed at $p=-\rho$, finite junction conditions on the null horizon surface necessarily lead to an interior solution with a singular core, where the perturbative expansion breaks down. Comparison to other models and implications for more rapidly rotating gravastars are briefly discussed.
\end{abstract}

\begin{description}
\item[PACS numbers]
\end{description}
\maketitle

\section{Introduction}

Extension of the exterior geometry of a black hole (BH) through the horizon and into the interior involves the physical assumption that the vacuum Einstein equations apply at and inside the horizon. Motivation for considering alternative BH interiors arises first from the curvature singularities in the geometries 
of mathematical BHs, which according to the classical singularity theorems are a general feature once some set of conditions involving trapped surfaces and energy conditions applies \cite{Penrose:1965,HawkPen:1970}. In addition to curvature singularities, the BH interiors may show acausal characteristics, such as 
the closed timelike curves of the Kerr solution~\cite{HawkingEllis:1973}. 
Such acausal behavior is widely held to be unphysical, especially since it arises on macroscopic scales comparable to the scale of the BH horizon.

 A BH horizon is a marginally trapped surface from which matter and information cannot exit classically. This horizon 
boundary is the root of severe difficulties BHs pose for quantum theory, most notably the apparent nonconservation of probability and enormous BH entropy implied by the Hawking effect \cite{Hawking:1976}. The resulting ``BH information paradox" \cite{tHooft:1995} has been the source of perplexity and speculation 
for more than four decades, with a wide range of views on its possible resolution \cite{UnruhWald:2017}, some of them quite radical, in which no classical spacetime 
interior may survive at all \cite{Mathur:2015}. 

Another noteworthy property of horizons is the potentially large, unbounded semiclassical stress tensor from vacuum polarization effects that can
occur on them \cite{ChristenFulling:1977,MottolaVaulin:2006,EMZak2010}. If by any mechanism a nonvanishing surface stress tensor is present on the horizon, a globally vacuum solution is inappropriate and the interior may not be singular or have any unphysical or acausal features at all. 
Consistency with quantum unitary evolution can then be maintained in this case. Physical surface stresses on a BH horizon, whatever their origin, 
are perfectly consistent with the equivalence principle as any physical surface is, and can be described within classical general relativity in terms of a 
sharply peaked anisotropic stress tensor---in the infinitely thin surface limit, a Dirac $\d$-function distribution---localized on the horizon. 
 
An explicit example of surface stresses and a quite different interior solution is provided by the gravitational condensate star (gravastar) in \cite{MazurMottola:2015}, where the Schwarzschild BH exterior is matched to a non-singular de Sitter interior with a positive 
vacuum energy but negative pressure, $p\!=\! -\r$, at a surface located at their respective horizons. A similar proposal based on an analogy to quantum phase transitions in condensed matter was suggested in Ref.~\cite{ChapHohlLaughSant:2001}. An abrupt change in ground
state vacuum energy at the horizon is characteristic of a quantum phase transition, and would lead to a dark energy de Sitter interior if $\r = -p > 0$~\cite{EMZak2010}. Independently 
of the microscopic origin of the phase transition boundary layer, the interior static de Sitter and exterior Schwarzschild geometries are ``glued" together at their respective horizons at $\RS = 2GM/c^2=c/H$, where $M$ is the Schwarzschild mass and $H$ is related to the de Sitter energy density by $\r = 3H^2/(8\pi G)$. The discontinuity $[\k]$ in the surface gravities at the horizon results in a positive surface 
tension $\t_s = [\k]/8 \pi G = c^4(8\pi G \RS)^{-1}$ \cite{MazurMottola:2015}, and a surface stress tensor $\d$-function localized on the gravastar physical surface that replaces the mathematical BH horizon.

A variety of other models for ``non-singular BHs" or ``BH mimickers" have been proposed over the years~\cite{PoisIsrae:1988,Frolov:1989,Dymnikova:1992,Dymnikova:2003,Ansoldi:2008jw}. 
In addition to their different interiors, several of these models have also been called  ``gravastars" by their authors, some with thin shells~\cite{Visser:2003ge,Pani:2015,Uchikata:2015,UchiPani:2016}, some with shells of quite large macroscopic thickness consisting of stiff matter~\cite{ChirentiRezz:2007,ChirentiRess:2016} or anisotropic matter~\cite{CattVisser:2005,ChirentiRezz:2007,ChirentiRezz:2008}. It is important to recognize that these alternate models have a {\it timelike} outer boundary at radius
$R > 2GM/c^2$, {\it not} at the null horizon as in \cite{MazurMottola:2015}. This distinction is important
 in whether or not the usual Israel junction conditions \cite{Israel:1966a} can be used, since these are unsuitable for null hypersurfaces \cite{BarrabesIsrael:1991}.

More importantly, the distinction between timelike and horizon surfaces manifests itself in physical properties and observational signatures that can distinguish such ultracompact objects from mathematical BHs, for example in their tidal deformability. In the era of gravitational wave and multimessenger astronomy it becomes of paramount importance to distinguish these various alternative models, with
different consequences for observations.

At the present time there is 
no direct observational evidence of the existence of a BH event horizon. Indirect arguments, based upon accretion models \cite{BrodNary:2007,BrodLoebNary:2009}, are limited by the assumptions of the 
models  \cite{CarbDiFiLibViss:2018}.  The LIGO/LSC gravitational wave data~\cite{LIGO:2020}, and even the spectacular 
images of M87 obtained by the Event Horizon Telescope \cite{EHT1435168}, are sensitive to the light rings at the innermost stable circular orbits, probing the geometry well outside the BH horizon~\cite{CardFranPani:2016,CardosoPani:2019}.  As a result, while some thick shell variants are already ruled out \cite{ChirentiRess:2016},
the gravastar of Ref.~\cite{MazurMottola:2015}, defined by an extremely thin shell localized on the horizon, remains consistent with observations \cite{CardosoPani:2019}. The physics of a surface at or near a horizon, and possible interior structure of ultracompact thin-shell models, can be probed by the after-merger
ringdown and/or new phenomena such as ``echoes" of the original merger event signal \cite{CardosoPani:2019,CardHopMacPalPani:2016}, 
for which at present, there are suggestions but no conclusive evidence \cite{AbediDykAfsh:2017,NielCapBirnWester:2019}.  

In this paper we shall reserve the term ``gravastar" for the universal gravitational condensate 
star solution of Ref.~\cite{MazurMottola:2015}, where the lightlike null horizon plays a privileged role as the locus of joining of interior and exterior classical geometries, and the surface layer there is of negligible thickness, so that its stress tensor is well approximated by a Dirac $\d$-function.
Its generalization to the case of slow rotation is the subject of this paper.

Given the spherically symmetric gravastar, the method of Hartle-Thorne \cite{Hartle:1968si,Chandrasekhar:1974} can be used to find slowly rotating solutions by expanding 
in powers of a small rotation parameter, under the assumption that the equation of state of the interior is unchanged from the nonrotating case. In Ref.~\cite{Posada:2016} this method was applied to a slowly rotating constant-density Schwarzschild star of mass $M$ and its radius $R>2GM/c^2$ was allowed to approach $2GM/c^2$ numerically. In this paper we show that analytic solution of the perturbation equations to second order in the rotation and matching directly on the 
horizon at radius $2GM/c^2$ is possible by the newly developed methods of \cite{axihorizon}, hereafter referred to as Paper I. As we shall see, matching at the horizon leads to different 
conclusions from that of~\cite{Posada:2016}, or that of other models matched at a radius different from $2GM/c^2$~\cite{Pani:2015,Uchikata:2015,UchiPani:2016}.

The structure of the paper is as follows. In the next section the nonrotating spherical gravastar of Ref.~\cite{MazurMottola:2015} is briefly reviewed. In Sec.~\ref{Sec:Eineqs}
we derive the Einstein equations for the slowly rotating gravastar.
In Secs.~\ref{Sec:Exterior} and \ref{Sec:Interior} we give the general analytic solution of these equations for the exterior and interior rotating gravastar metric functions respectively,
showing that finiteness of the perturbations on the horizon requires the exterior geometry to be identical to that of a Kerr BH.
In Sec.~\ref{Sec:Joining}, we discuss the joining of the interior and exterior at the mutual horizon boundary, relying on the analytic method of Paper I \cite{axihorizon}. In 
Sec.~\ref{Sec:SurfTens} we give the $\d$-distributional stress energy tensor on the null horizon and surface gravity discontinuity resulting from this gluing of interior to exterior.
Section~\ref{Sec:Komar} contains the Komar mass, angular momentum and moment of inertia of the slowly rotating gravastar solution, showing that some integration constants
can be fixed by eliminating sources at the origin.  Section~\ref{Sec:SumDisc} contains a Summary and Discussion of our
results, and comparison to other models, while Sec.~\ref{Sec:Conclusions} contains our Conclusions and consequences for observations.

There are three Appendices. Appendix \ref{Sec:EinTens} contains the Einstein tensor component for the rotating metric (\ref{axisymstat}) expanded up to
second order in the small rotation parameter. Appendix \ref{Sec:Weyl} contains the Weyl tensor
of this metric, and Appendix \ref{Sec:Conform} contains the conformal diagram that results from gluing of the nonrotating gravastar interior to the exterior at the horizon surface.
Hereafter we generally use geometric units where $G\!=\!c\!=\!1$ to simplify the notation, restoring them only when useful for clarity, and MTW metric and curvature conventions \cite{MTW}.

\section{The Spherical Gravastar and Horizon Surface Tension}
\label{Sec:NRotat}

In order to establish notation and conventions we first briefly review the spherically symmetric nonrotating gravastar of Ref.~\cite{MazurMottola:2015}. As shown in~\cite{MazurMottola:2015}, this gravastar solution can 
be obtained from Schwarzschild's constant density interior solution \cite{Schwarz:1916,Synge:1960} by a limiting process in which the star radius $R\to \RS \equiv 2GM/c^2$,
the Schwarzschild horizon radius. 
The gravastar of~\cite{MazurMottola:2015} may be viewed as a universal limit of the gravitational condensate star
model first proposed in \cite{MazurMottola:2001,MazurMottola:2004}, where a thin layer of ultrarelativistic $p=\rho$ material was interposed, straddling the mutual Schwarzschild 
and de Sitter horizons. It is universal in the sense that when the thickness of the intervening surface layer localized on the horizon is taken to zero, the resulting solution 
is independent of any assumptions about the intervening layer equation of state, and the surface stress tensor is instead completely determined by the matching of the interior and
exterior spacetimes on their respective horizons.  This universal (nonrotating) gravastar has maximal compactness $GM/R = 1/2 $ (up to possible Planck scale corrections),  and evades the 
Buchdahl bound $R \ge (9/4) GM$, applicable to isotropic fluid spheres \cite{Buchdahl:1959}, by having an anisotropic stress at its surface. 
 Like the arbitrarily thin shell 
model of \cite{MazurMottola:2001,MazurMottola:2004}, the gravastar of \cite{MazurMottola:2015} is a low entropy, cold condensed solution, in no conflict with unitary quantum evolution or statistical mechanics, possesses
no enormous entropy, and hence no information paradox \cite{MazurMottola:2001,MazurMottola:2004}.

The general static, spherically symmetric line element can be expressed as
\begin{align}
ds^2 = -f(r)\, dt^2 + \frac{dr^2}{h(r)} + r^2 \big(d\th^2 + \sin^2\!\th \, d\f^2\big)
\label{sphstat}
\end{align}
with $f(r), h(r)$ two arbitrary functions of $r$.  Alternately one may write
\begin{align}
f(r) = e^{2\n_0(r)} = \frac{h(r)}{[j(r)]^2}, \qquad h(r) = 1 - \frac{2m(r)}{r} 
\label{sphstat2}
\end{align}
in terms of the gravitational potential $\n_0 (r)$, the mass function $m(r)$, and Hartle's function $j(r)$, the latter defined to be positive.

The general spherically symmetric solution of Einstein's equations requires the three stress-energy tensor components 
$T^t_{\ t} = - \r$, $T^r_{\ r} = p_r$, and $T^\th_{\ \th} = T^\f_{\ \f}= p_{\perp}$, which are functions of $r$ only. If $p_{\perp} = p_r \equiv p$, the pressure is isotropic. If in addition $\r + p = 0$, as in the exterior and interior of a gravastar, Einstein's equations imply that $j(r)$ is constant and $m(r)$ is either a Schwarzschild term or a de Sitter term (or a sum of the two).

The spherical nonrotating gravastar solution is given by the following piecewise continuous functions $f(r)$ and $h(r)$ in the condensate interior and vacuum exterior respectively~\cite{MazurMottola:2015},
\begin{subequations}
\label{gravNrot}
\begin{align}
  f(r) &=
  \begin{cases}
    \displaystyle \dfrac{1}{4}\left(1- \sdfrac{r^2}{\RS^2}\right)\,,
    &  0\le r \le \RS
    \\[1ex]
    \displaystyle 1- \sdfrac{\RS}{r}\,,
    & r \ge \RS
  \end{cases}
  \\
  h(r) &=
  \begin{cases}
    \displaystyle 1- \sdfrac{r^2}{\RS^2}\,,
    & 0\le r \le \RS
    \\[1ex]
    \displaystyle 1- \sdfrac{\RS}{r}\,,
    & r \ge \RS\,.
  \end{cases}
\end{align}
\end{subequations}
The line element (\ref{sphstat}) with (\ref{gravNrot}) describes an interior static patch of de Sitter space with  
\begin{align}
 \r = - p =\, \frac{3H^2\!\!}{\!8\pi} \,= \frac{3}{8\pi \RS^2}\,, \qquad r< \RS 
\label{rhop}
\end{align}
joined to a Schwarzschild exterior with 
\begin{align}
 \r = p =0\,, \qquad  r > \RS
\label{rhop0}
\end{align}
at their mutual horizons
\vspace{-3mm}
\begin{align}
r=\RS  = 2M =H^{-1}\,.
\label{RH}
\end{align}
The function $j(r)$ 
\vspace{-3mm}
\begin{align}
  j(r) =  \sqrt{\sdfrac{h}{f} } &=
  \begin{cases}
    2\,,&  0\le r < \RS \\
    1\,,&  r> \RS
  \end{cases}
\label{jdef}
\end{align}
is discontinuous at the horizon surface, and the mass function is given by
\begin{align}
m(r) = 
\begin{cases}
  \displaystyle \sdfrac{1}{2} H^2r^3\,,
  &0\le r < \RS
  \\ 
  \displaystyle M\,,
  &r > \RS\,.
\end{cases}
\end{align}
Note that just as a Schwarzschild BH can have any mass, so too the spherically
symmetric gravastar can have any mass $M$, provided only that the de Sitter vacuum dark energy density is fixed by (\ref{rhop}) and (\ref{RH}).

The discontinuity $[\k]= \k_+ - \k_-$ in the surface gravity
\begin{align}
\k(r) = \frac{1}{2} \sqrt{\sdfrac{h}{f}}\, \frac{df}{dr} &=
  \begin{cases}
      \displaystyle -\frac{1}{2\RS}\,,&  r \to \RS^- \\[2ex]
      \displaystyle + \frac{1}{2\RS}\,,&  r\to \RS^+
  \end{cases}
\label{surfgrav}
\end{align}
as $r\to \RS^{\pm}$, gives rise to a $\d$-distributional anisotropic stress tensor whose nonzero components are
\begin{align}
^{(\S )}T^A_{\ \,B}\, \sqrt{\sdfrac{f}{h}} = \cS^A_{\ \,B} \, \d(r-\RS)\,,\qquad  \cS^A_{\ \,B}= \frac{[\k]}{8\pi G} \, \d^A_{\ \, B} 
\label{surfTsph}
\end{align}
for $A, B = \th, \f$. This stress tensor corresponds to a surface tension \cite{MazurMottola:2015,axihorizon}
\begin{align}
\t_s =  \frac{[\k]}{8\pi G} = \frac{c^4}{8\pi G \RS} =  \frac{c^6}{16\pi G^2 M}\,.
\label{taunonrot}
\end{align}
The surface stress tensor $^{(\S )}T^A_{\ \, B}$ in (\ref{surfTsph}) is a well-defined distribution when integrated against the standard coordinate invariant volume measure 
$d^4x \sqrt{-g} = dt\,dr\, r^2 \sqrt{f/h}\ d\th \,d\f\sin\th$, since $r^2 \sin\th$ is continuous on the horizon. The result (\ref{surfTsph})-(\ref{taunonrot}) 
may be derived directly from the Einstein tensor density for the piecewise continuous metric (\ref{gravNrot}) \cite{axihorizon}.

The surface stress $\cS^A_{\ \,B}$ 
differs from that obtained from the usual Israel junction conditions by  \cite{MazurMottola:2015,axihorizon}
\vspace{-4mm}
\begin{align}
\big(\cS^A_{\ \,B}\big)_{\rm here}  = \sqrt{f (R)} \, \big(S^A_{\ \,B}\big)_{\rm Israel} \,.
\label{Isrmod}
\end{align}
Whereas the junction conditions as originally formulated in~\cite{Israel:1966a} do not apply to the null horizon hypersurface where $f(\RH) \!=\! 0$, junction conditions modified by the $\sqrt{f}\!=\!e^{\n_0}$ redshift factor and
with one contravariant and one covariant index as in (\ref{Isrmod}) give the well-defined finite result (\ref{surfTsph})-(\ref{taunonrot}) \cite{axihorizon}. The relation (\ref{Isrmod})
is important for any comparison of the gravastar with its horizon surface tensor (\ref{surfTsph}) to other models with matching at some radius away from the horizon
by Israel's method, in either the cases of zero or nonzero angular momentum.

\section{Linearized Einstein Equations for Slow Rotation}
\label{Sec:Eineqs}

We are interested in small stationary and axisymmetric perturbations due to rotation
about the static spherically symmetric gravastar solution summarized in the previous section.
There are several forms for the general axisymmetric and stationary metric line element in the literature which may be used for this purpose
\cite{Lewis:1932,Papapetrou:1966,HartleSharp:1967,Hartle:1967he,Bardeen:1970,FriedmanChandra:1972,Chandrasekhar:1984}. The condition of stationarity and axial symmetry is invariantly defined by the geometry through two independent Killing vectors corresponding to time translations and rotations around the fixed rotation axis. Coordinates $t$ and $\phi$ adapted to these symmetries can be introduced such that these Killing vectors are ${\pa}/{\pa t} $ and $ {\pa}/{\pa \f}$, respectively. The two remaining coordinates $r$ and $\theta$ can be chosen so that the stationary axisymmetric line element assumes the form
\begin{align}
ds^2 = -e^{2\n} dt^2 + e^{2\j} \big(d \f - \w\, dt\big)^2 + e^{2 \a} dr^2 + e^{2 \b} d\th^2
\label{axisymstat}
\end{align}
where the five functions $\n, \j, \a, \b, \w$ are functions of $r$ and $\theta$.  There remains the freedom to define a coordinate condition to reduce these five functions to the minimal four functions necessary to specify the general such metric. Hartle's condition  \cite{Hartle:1967he}
\begin{align}
e^\j = e^\b \sin\th
\end{align}
fixes this local coordinate freedom, leaving reparametrizations of $r$ still allowed.

Consider now a stationary, axisymmetric, slowly rotating, nearly spherically symmetric solution, {\it i.e.} small perturbations about the static solution that are invariant under the transformations $(t,\phi)\to(-t,-\phi)$ and $\theta\to\pi-\theta$. Hartle and Chandrasekhar-Miller expanded the line element (\ref{axisymstat}) to second order in the angular momentum as~\cite{Hartle:1967he,Chandrasekhar:1974}
\begin{align}
ds^2 = & - e^{2\n_0(r)} \Big[ 1+2 h_0(r) + 2 h_2(r)\, P_2(\cos\th)  \Big]  dt^2 \nn & 
+ \frac{r}{r-2m(r)}  \left\{1+\frac{2}{r-2m(r)} \Big[ m_0(r) + m_2(r)\, P_2(\cos\th) \Big] \right\}  dr^2 \nn& 
+ r^2 \Big[ 1+2 k_2(r)\, P_2(\cos\th)  \Big] \Big[ d\th^2 + \sin^2\!\th\,\big(d\f - \w(r) dt \big)^2 \Big].
\label{kingmet}
\end{align}
Here $P_l(\cos\th)$ is the Legendre polynomial of order $l$, $m(r)$ and $\nu_0(r)$ are the metric functions of the nonrotating solution, and $h_l(r)$, $m_l(r)$, $k_l(r)$ are the monopole ($l=0$) and quadrupole ($l=2$) contributions of second order in rotation respectively. The choice $k_0(r)=0$ is part of Hartle's choice of gauge.
The function $\w(r)$ is the first-order contribution that gives rise to inertial frame dragging. 

The metric line element (\ref{kingmet}) is an expansion of (\ref{axisymstat}) with the identifications
\begin{subequations}
\begin{align}
&e^{2\n} =  e^{2\n_0(r)} \Big[ 1 + 2 h_0(r) + 2 h_2(r) \,P_2(\cos \th)\Big] \label{eq:exp2nu} \\
&e^{2\j} =  r^2 \sin^2\th\,  \Big[1  + 2 k_2(r)\, P_2(\cos \th)\Big]\\
&e^{2\a} =  \frac{r}{r-2 m(r)} \left[1 +  2 \frac{m_0(r)+m_2(r)\,P_2(\cos \th)}{r- 2 m(r)}\right]\\
&e^{2\b} =  r^2 \Big[1  + 2 k_2(r) \, P_2(\cos \th)\Big]
\end{align}
\end{subequations}
which will be used both in the interior and exterior of the slowly rotating gravastar, matching at the gravastar null horizon surface.

The perturbed stress-energy tensor is taken to be of the same form as in the nonrotating background gravastar solution, namely
\vspace{-4mm}
\begin{align}
T^\m_{\ \,\n}  =  -\EE \,\d^\m_{\ \n}
\label{EMT}
\end{align}
where $\EE$ can be expanded up to second order in the rotation as
\begin{align}
& \EE = \r(r) + \EE_0(r) + \EE_2(r) P_2(\cos\th) \,.
\end{align}
Here $\r$ is the background energy density, constant in the interior and exterior of the gravastar, given by (\ref{rhop})-(\ref{rhop0}),
and $\EE_0$, $\EE_2$ are its monopole and quadrupole perturbations. The covariant conservation equation $\na_\m T^\m_{\ \,\n} = 0$ for the $T^\m_{\ \,\n}$ in Eq.~\eqref{EMT} becomes $\pa_\m \EE=0$,  and in particular its $\n=r,\th$ components and $\rho={\rm const}$ require 
\vspace{-3mm}
\begin{align}
\label{eq:E0_E2}
\EE_0 = {\rm const}, \qquad \EE_2= 0 
\end{align}
in the interior and exterior regions, with $\EE_0=0$ in the exterior.

The forms of the perturbed Einstein equations in Refs.~\cite{Hartle:1967he,Chandrasekhar:1974} cannot be applied directly to the case of a rotating gravastar, because they were derived 
assuming a perfect fluid under uniform rotation $\W$, and use definitions of pressure and energy density perturbations that become undefined when $\r+p=0$, which is the
case of interest here. We have therefore derived the perturbed Einstein equations for the cases of a slowly rotating gravastar anew. The components of the Einstein tensor for the metric (\ref{kingmet}) are given in Appendix~\ref{Sec:EinTens}.

Because $j$ is constant in both the interior and exterior of the gravastar, and thus its derivatives vanish there, and because of the specific constant form 
of the energy-momentum tensor (\ref{EMT}), having the same $p=-\r$ equation of state as the nonrotating case, the resulting linear perturbation equations are somewhat simpler than the perfect 
fluid case considered in~\cite{Hartle:1967he,Chandrasekhar:1974}. In particular, the uniform angular velocity $\W$ and the auxiliary function $\varpi=\W-\w$ of~\cite{Hartle:1967he,Chandrasekhar:1974} do not appear.

The first order perturbed Einstein equation
\begin{subequations}
\label{mastereqns}
\begin{align}
&\frac{d}{dr}\left(r^4\frac{d \w}{dr}\right)=0
\label{masterw}
\end{align}
comes from $G^t_{\ \f}=0$ and gives the frame dragging function $\w(r)$. Making use of (\ref{masterw}), the
second order monopole terms in the Einstein equations $G^t_{~t}=8\pi T^t_{~t}$ and $G^r_{~r}=8\pi T^r_{~r}$ are
\begin{align}
& \frac{dm_0}{dr} = 4 \pi r^2 \EE_0 + \frac{j^2 r^4}{12}  \bigg( \frac{d\w}{dr} \bigg)^2  
\label{masterm0}
\\
& \frac{dh_0}{dr} = \frac{1-2m'}{(r-2m)^2} \, m_0 - \frac{1}{r-2m} \frac{dm_0}{dr}
\label{masterh0}
\end{align}
Here $m' \equiv dm/dr=4\pi r^2\r$. These equations can be combined to give 
\begin{align}
 h_0 = - \frac{m_0}{r-2m} + C
\label{masterh0m0}
\end{align}
where $C$ is a constant.
For the quadrupole functions, with $v_2=h_2+k_2$ from \cite{Chandrasekhar:1974}, we find
\begin{align}
&\hspace{4cm}m_2  = - (r-2m) \bigg[ h_2  - \frac{1}{6} j^2 r^4 \bigg( \frac{d\w}{dr} \bigg)^2 \bigg] \label{m2xx}
\\
&\hspace{4cm}\frac{dv_2}{dr} = - 2 \nu_0' \, h_2 + (1 + r \nu_0') \, \frac{j^2r^3}{6}  \bigg( \frac{d\w}{dr} \bigg)^2 \label{v2xx}
\\
&\frac{dh_2}{dr} = - \frac{2}{r(r-2m) \nu_0'} \, v_2 
- 2 \bigg(  \nu_0'   + \frac{m}{r^2(r-2m) \nu_0'}  \bigg)   h_2  
+ \bigg( \nu_0'- \frac{1}{2r(r-2m) \nu_0'}\bigg) \frac{j^2r^4}{6} \bigg( \frac{d\w}{dr} \bigg)^2  
\label{h2xx}
\end{align}
\end{subequations}
where $\nu_0' \equiv d\nu_0/dr=(m-4\pi r^3 \rho)/[r(r-2m)]$.
The algebraic equation for $m_2$ comes from the quadrupole perturbations of $G^\theta_{~\theta}-G^\phi_{~\phi}=0$. The equation for $v_2$ comes from the equation $G^r_{~\theta}=0$ with the replacement of $m_2$ from Eq.~(\ref{m2xx}). Finally, the differential equation (\ref{h2xx}) for $h_2$ comes from the quadrupole perturbation term in $G^r_{~r}=8\pi T^r_{~r}$. Eq.~\eqref{m2xx} can also be written in the form
\begin{align}
h_2 = - \frac{m_2}{r-2m} + \frac{1}{6} j^2 r^4 \bigg( \frac{d\w}{dr} \bigg)^2
\label{masterh2m2}
\end{align}
which is analogous to~\eqref{masterh0m0}. The function $k_2$ is then given by $k_2= v_2-h_2$.

\section{Exterior solution}
\label{Sec:Exterior}

Outside the gravastar ($r>2M$), the spacetime is empty, and from (\ref{gravNrot})-(\ref{rhop}), we have
\begin{align}
&m(r)=M\,,\qquad h=f = e^{2\n_0} = 1 - \frac{2M}{r}\,,\qquad j=1\,.
\end{align}
The solution of (\ref{masterw}) for the frame dragging that goes to zero as $r\to \infty$ is
\begin{align}
\w = \frac{2J}{\,r^3\!}
\label{boext}
\end{align}
where $J$ is an integration constant that can be identified as the total angular momentum of the solution.
Using \eqref{boext} in the Eq.~\eqref{masterm0} for $m_0$, and then using~\eqref{masterh0m0} for $h_0$, gives
\begin{subequations}
\label{extmonofns}
\begin{align}
& m_0 = \d M - \frac{J^2}{r^3}\\
& h_0 = - \frac{m_0}{r-2M} + C_{_E}  \label{h0Cext} 
\end{align}
\end{subequations}
in terms of two integration constants $\d M$ and $C_{_E}$ in the exterior.

For the quadrupole functions $m_2,k_2,h_2$ in the exterior, 
the two coupled linear first order equations~(\ref{v2xx}) and~(\ref{h2xx}) for $v_2$ and $h_2$ can be converted to a single second 
order equation for $h_2$, which with the change of variable to $z= r/M -1$ is recognized as the associated Legendre differential equation of degree $l=2$ and order $m=2$, with a particular inhomogeneous term
in $\w^{\prime\, 2}$. The general solution for $h_2$ is thus the sum of a particular solution of the inhomogeneous equation and two independent solutions of the 
associated Legendre equation, which can be taken to be the associated Legendre functions of the first and second kind, namely $P^2_2(z)=3(z^2-1)$ and
\begin{align}
Q_2^2(z) = \frac{3(z^2-1)}{2} \ln\left(\frac{z+1}{z-1}\right)  + \frac{z(5-3z^2)}{z^2-1}\,. \label{Q22}
\end{align} 
The branch cut in this function is chosen along the real axis from $-\infty$ to $+1$, since in the exterior of the gravastar, $r>2M$ and thus $z=r/M-1>1$.
The function $v_2$ then follows from $h_2$ by use of Eq.~\eqref{v2xx}, and contains associated Legendre functions  $P_2^1(z)=3z(z^2-1)^{1/2}$ and $Q_2^1(z)$, again with a branch along the real axis from $-\infty$ to $+1$, where
\begin{align}
\frac{1} {\sqrt{z^2 -1}}\, Q_2^1(z) = \frac{3z}{2}\ln\left(\frac{z+1}{z-1}\right) + \frac{\ 2-3z^2\!}{\!z^2-1\!\!}\,. \label{Q21}
\end{align} 
The general solutions for the quadrupole perturbations are thus found to be
\begin{subequations}
\label{extquadfns}
\begin{align}
& h_2 = \frac{J^2(r+M)}{Mr^4}+ A_{_E} \, \frac{r(r-2M)}{2M^2}  + B_{_E} \, Q_2^2\!\left(\sdfrac{r}{M}-1\right) \label{h2ext}\\
& v_2 = h_2 + k_2 = -\frac{J^2}{r^4} - A_{_E}\, \frac{r-M}{M} - \frac{2 B_{_E} M}{\sqrt{r(r-2M)}} \, Q_2^1\!\left(\sdfrac{r}{M}-1\right)\label{v2ext} \\
& m_2= - (r-2M) \left( h_2  - \frac{6J^2}{r^4} \right)  
\end{align}
\end{subequations}
with $A_{_E}$ and $B_{_E}$ two additional constants of integration besides $J$, $\d M$, and $C_{_E}$.

If we require the solution to be asymptotically flat, then $A_{_E}=0$, and moreover a nonzero $C_{_E}$ in (\ref{h0Cext}) is simply a rescaling of the time coordinate $g_{tt}(A_{_E}=0, r\to\infty)=-1-2C_{_E}$.
If the time coordinate $t$ and Killing vector $\pa/\pa t$ are fixed by $g_{tt} \to -1$ asymptotically, then $C_{_E}=0$ is also fixed. The asymptotically flat exterior
Hartle-Thorne metric (\ref{kingmet}) depends then on the remaining constants $J$, $B_{_E}$, and $\d M$. 
As $r \to 2M$, the quadrupole function $k_2$ behaves as
	\begin{align}
	k_2 & \to  B_{_E} \left( 8 + 3 \ln\frac{r-2M}{2M} \right) - \frac{J^2}{4M^4}
\label{eq:k2H_ext}
	\end{align}
where we retain the subdominant terms for later reference. Not only is $k_2$ supposed to be a small perturbation, but when it diverges the equatorial circumference also diverges.
Finiteness of $k_2$ at $r=2M$ requires
\vspace{-4mm}
\begin{align}
B_E=0.
\label{BEfix}
\end{align}
Note that $B_E=0$ also eliminates the $B_E/(r-2M)$ behavior in the function $h_2$ as $r\rightarrow 2M$.
 With \eqref{BEfix}, the exterior solution simplifies to 
\begin{subequations}
\label{extfinal}
\begin{align}
& m_0 =  \d M - \frac{J^2}{r^3} \\
& h_0 = - \frac{m_0}{r-2M} \\
& h_2 = \frac{J^2}{r^3} \bigg( \frac{1}{r}  + \frac{1}{M}\, \bigg)  \label{extfinal_h2}\\
& m_2 =  \frac{J^2(r-2M)(5M-r)}{Mr^4} \\
& k_2 = -\frac{J^2(r+2M)}{Mr^4}\\
&\  \w=\frac{2J}{r^3}\,.
\end{align}
\end{subequations}
With \eqref{extfinal}, we see that the combined term $e^{2\nu_0} h_0 = -m_0/r$ in the metric function $e^{2\nu}$ in \eqref{eq:exp2nu} remains finite at all $r \ge 2M$, while the function $h_0$ itself generally diverges
\begin{align}
h_0 \to - \frac{1}{r-2M} \bigg(  \d M - \frac{J^2}{8M^3}\bigg)
\end{align}
as $r \to 2M$ unless
\begin{align}
\d M = \frac{J^2}{\RS^3} = \frac{J^2}{8M^3}\,.
\label{eq:dRH_ext}
\end{align}

As shown originally in Ref.~\cite{Hartle:1968si}, the Hartle-Thorne metric with $A_{_E}\!=\!B_{_E}\!=\!0$ is equivalent to the Kerr metric for a BH of total mass $\MM=M+\d M$ 
and angular momentum  $J=a\MM \simeq a M$ to second order in rotation. In fact, the Kerr metric  to second order in $a\!=\!J/\MM$ in Boyer-Lindquist coordinates $(r_{_{BL}} ,\th_{_{BL}} )$ results from the Hartle-Thorne metric in (\ref{kingmet}) with~(\ref{extfinal}) through the coordinate transformation 
\begin{subequations}
\label{BL2HT}
\begin{align}
& r_{_{BL}} =  r \bigg\{ 1 - \frac{a^2}{2r^2} \bigg[ \bigg( 1 + \frac{2\MM}{r} \bigg) 
\bigg( 1 - \frac{\MM}{r} \bigg) - \cos^2\theta \bigg( 1 - \frac{2\MM}{r} \bigg) \bigg( 1 + \frac{3\MM}{r} \bigg) \bigg] \bigg\} ,\\
& \th_{_{BL}} = \th - \frac{a^2}{2r^2} \cos\th \sin\th  \bigg( 1 + \frac{2\MM}{r} \bigg) 
\end{align}
\end{subequations}
where we have corrected the sign of the $\cos^2\!\th$ term in~\cite{Hartle:1968si}.

Thus the requirement that the perturbations of the spherically symmetric gravastar solution due to rotation remain finite on the gravastar horizon surface, 
together with asymptotic flatness, forces the exterior geometry to be {\it identical} to the second order expansion of a slowly rotating Kerr BH. A slowly rotating gravastar has compactness 
$M/\RS = 1/2$ (up to possible Planck scale corrections), and cannot be distinguished from a Kerr BH by any measurement or observation sensitive only 
to the external spacetime metric. Note that this conclusion would not follow if we did not require finiteness of the perturbations on the horizon, since (\ref{BEfix}) would not be required in that case.

\section{Interior Solution}
\label{Sec:Interior}
The background metric of the gravastar interior is a region of de Sitter space with a rescaled time coordinate, 
with
\begin{align}
m(r)=\sdfrac{1}{2}\, H^2r^3\,,\qquad h = 1-H^2r^2\,,\qquad j=2\,.
\end{align}
We temporarily leave $j$ unspecified to allow for easier comparison to other work \cite{UchiYoshida:2014,Uchikata:2015,Pani:2015,UchiPani:2016}, where typically $j=1$ is used for Hartle perturbations on de Sitter space.
The frame dragging and monopole perturbations are again found by integrating (\ref{masterw})--(\ref{masterh0m0}),
\begin{subequations}
\label{int_soln}
\begin{align}
& \w =W_1+\frac{W_2}{r^3} \label{int_soln_a}\\
& m_0 = \d M_{I}  + \frac{4 \pi  \EE_0}{3} r^3-\frac{j^2 W_2^2}{4 r^3}\\
& h_0 = - \frac{m_0}{r(1-H^2r^2)} +C \label{int_soln_h0} 
\end{align}  
\end{subequations}
where $W_1,W_2,\d M_I, C$ are integration constants. The stress energy tensor for the interior is given by (\ref{EMT})--(\ref{eq:E0_E2}) with 
(\ref{rhop}) and nonzero $\EE_0$ in general.
   
The quadrupole perturbations are a linear combination of two independent homogeneous solutions and a particular solution of Eqs.~(\ref{v2xx})--(\ref{h2xx}). Upon making the change of variable to $z = 1/(Hr)$, the homogeneous equations may again be combined into an associated Legendre differential equation of degree $l=2$ and order $m=2$, this time for $rh_2$, whose solutions are the associated Legendre functions
$P^2_2(z)$ and $Q^2_2(z)$ of (\ref{Q22}). The general solution for the quadrupole perturbations in the interior is then
\begin{subequations}
\label{int_soln2}
\begin{align}
v_2 &=h_2 + k_2 = -\frac{j^2 W_2^2 \left(3 -10 H^2 r^2 + 5H^4r^4\right)}{8 r^4} + \frac{2 A_I}{r} - \frac{B_I Hr}{\sqrt{1-H^2r^2}} \, Q_2^1(1/Hr) \label{int_soln_v}\\
m_2 & = \frac{j^2 W_2^2\left(1- H^2r^2\right)  \left(8 - 5 H^2r^2\right)}{4r^3} + A_I\frac{(1-H^2r^2)^2}{H^2r^2} -\frac{B_I(1- H^2r^2)}{2H} \, Q_2^2(1/Hr) \label{m2int}\\
h_2 &= - \frac{m_2}{r(1-H^2r^2)} + \frac{3j^2 W_2^2}{2r^4}  \label{int_soln_d} 
\end{align}
\end{subequations}
which together with \eqref{int_soln} depends upon the six integration constants $W_1$, $W_2$, $\d M_{I}$, $A_I$, $B_I$, and $C_I \equiv C$.

As $r\to 1/H$, the perturbation functions $m_0$, $m_2$ are finite while $k_2$ behaves as
\begin{align}
k_2 & \simeq 2HA_I - \frac{1}{2} j^2 W_2^2 H^4 + \frac{B_I }{2} \left( 7  + 3 \ln\frac{1-Hr}{2} \right)
\label{eq:k2H_int}
\end{align}
where we retain the subdominant terms for later reference.  
Finiteness of $k_2$ at the de Sitter horizon $r=1/H$ requires
\vspace{-4mm}
\begin{align}
B_I=0\,.
\label{BIfix}
\end{align}
The \eqref{BIfix} finiteness condition on the perturbation $k_2$ also removes the $B_I/(1- Hr)$ behavior from $h_2$.
From (\ref{int_soln_h0}), the combined term $e^{2\nu_0}h_0=-m_0/r$ in $e^{2\nu}$ in \eqref{eq:exp2nu} remains finite as $r\to1/H$, while the function $h_0$ itself generally diverges 
\begin{align}
h_0 & \to -\frac{H}{2(1-Hr)} \bigg( \d M_{I} + \frac{\!4\pi}{3H^3\!} \,\EE_0- \frac{j^2H^3W_2^2}{4}\bigg)
\end{align}
as $r \to 1/H$, unless
\vspace{-4mm}
\begin{align}
\EE_0 = \frac{3}{4\pi \RS^3} \left( \frac{j^2W_2^2}{4\RS^3} - \d M_I \right) \,.
\label{eq:elim_E0}
\end{align}

Finally, as $r\to0$, the functions $\w$, $m_0$, $m_2$, and $k_2$ diverge unless $W_2=A_I=0$. We do not impose these conditions since as we show in the following Sec.~\ref{Sec:Joining} it implies $J=0$ in the exterior and so no rotation at all of the gravastar, which is clearly too strong a condition. 

\section{Joining Exterior to Interior on the Null Horizon Surface}
\label{Sec:Joining}

The two Killing vectors $\pa/\pa t$, $\pa/\pa \f$ determine the location of the null hypersurface $r=\RH$ where the gluing of the interior and exterior solutions occurs
by the geometrically invariant condition that the norm of the vector $\ell = \pa/\pa t + \w \, \pa/\pa \f$ vanishes there, {\it i.e.}
\begin{align}
\ell \cdot \ell = e^{2\n} = f(r) \Big\{ 1 + 2h_0(r) + 2h_2(r) \,P_2(\cos \th)\Big\} = 0 \,.
\label{intH}
\end{align}
If we impose that $h_0$ and $h_2$ are finite at $r=2M$ in the exterior and at $r=1/H$ in the interior, resulting in Eqs.~\eqref{eq:dRH_ext} and~\eqref{eq:elim_E0}, and in addition use Eq.~\eqref{RH} from the nonrotating gravastar, then the solution of (\ref{intH}) is $r=\RH=\RS$.

In the coordinates $(t,\th,\f)$ adapted to the Killing vectors, the induced metric on the horizon hypersurface at $r=\RH$ is
\begin{align}
ds^2_H = \Sigma(\th) \, \Big[d\th^2 +  \sin^2\!\th\, \big(d\f  - \w_H \, d t \big)^2 \Big]
\label{surfmet}
\end{align}
where 
\vspace{-4mm}
\begin{align}
\w_H = \w(\RH) = \frac{2J}{\RH^3\!} 
\end{align}
and using~\eqref{eq:dRH_ext},~\eqref{eq:elim_E0},
\begin{align}
\S(\th) = e^{2\b(\RH,\,\th)} 
= \RH^2\Big[ 1 + 2 k_2(\RH) P_2(\cos\th) \Big]\,.
\label{Sigdef}
\end{align}
The induced metric~\eqref{surfmet} is regular on the null hypersurface, and independent of the perturbation functions $h_2$, $m_2$. The first junction condition is that the induced metric~\eqref{surfmet} is the same when approached from both sides, which requires that
$\RH$, $k_2(\RH)$, and $\w(\RH)$ have the same value on both sides. These conditions also ensure continuity of the inner products of the Killing vectors, \textit{i.e.}, $\pa/\pa t\cdot\pa/\pa t,~~\pa/\pa t\cdot\pa/\pa \phi,$ and $\pa/\pa \phi\cdot\pa/\pa \phi$.

The equality of the perturbation function $k_2(r)$ as $r \to \RH^{\pm}$ from both sides of the horizon implies that the interior limit
\vspace{-4mm}
\begin{align}
&\lim_{r\to \RH^-} k_2(r) = \frac{2A_I}{\RH} - \frac{j^2 W_2^2}{2\RH^4} 
\label{eq:k2H_int_lim}
\end{align}
from Eq.~\eqref{eq:k2H_int} with $B_I=0$ and $H=\RH^{-1}$, and the exterior limit
\begin{align}
&\lim_{r\to \RH^+} k_2(r) = - \frac{4J^2}{\RH^4}
\label{eq:k2H_ext_lim}
\end{align}
from Eq.~\eqref{eq:k2H_ext} with $B_E=0$ and $M=\RH/2$, coincide. Equating Eqs.~\eqref{eq:k2H_int_lim} and~\eqref{eq:k2H_ext_lim} gives
\begin{align}
A_I &= \frac{j^2W_2^2-8J^2}{4\RH^3}\label{AIfix}
\end{align}
fixing the interior integration constant $A_I$ in terms of $W_2$ and $J$. Finally, the equality of $\w(r)$ when the horizon is approached from both sides imposes
\begin{align}
W_1 = \frac{2J-W_2}{\RH^3} 
\label{W1fix}
\end{align}
and fixes the interior integration constant $W_1$.

With the finiteness conditions $B_E=B_I=0$, and the conditions \eqref{eq:dRH_ext}, \eqref{eq:elim_E0}, \eqref{AIfix}, \eqref{W1fix} on the horizon, the perturbation functions in the interior of the gravastar $r< \RH  = H^{-1} = 2M =\RS $, where $j=2$, become
\vspace{-3mm}
\begin{subequations}
\label{interiorsoln}
\begin{align}
\w&=\frac{2J-W_2}{\RH^3} + \frac{W_2}{r^3} \label{eq:int_omega} \\
m_0&= \delta M_I\left(1-\frac{r^3}{\RH^3}\right)+W_2^2\left(\frac{r^6-\RH^6}{r^3\RH^6}\right) \\
h_0&=C+ \frac{\RH^2m_0}{r(r^2-\RH^2)}  \label{eq:int_h0}\\
m_2&=\frac{W_2^2(r^2-\RH^2)}{r^3\RH^5}\left(r^3+5r^2\RH-r\RH^2-8\RH^3\right)-\frac{2J^2(r^2-\RH^2)^2}{r^2\RH^5} \label{eq:int_m2}\\
h_2&=\frac{6W_2^2}{r^4} + \frac{\RH^2m_2}{r(r^2-\RH^2)}\label{eq:int_h2}\\
k_2&=-\frac{W_2^2}{2r^4\RH^4}(r-\RH)(\RH^3+3r\RH^2+3r^2\RH+5r^3)-\frac{2J^2(r^2+\RH^2)}{r^3\RH^3}\label{eq:int_k2}
\end{align}
\end{subequations}
The interior solution \eqref{interiorsoln} still depends on the three interior integration constants:  $W_2$, $ \d M_I$, and $C$.  The first two of these can be fixed by the requirement of the absence of $\d$-functions at the origin, as we shall show in Sec.~\ref{Sec:Komar}. The Weyl curvature of the interior solution is given in Appendix \ref{Sec:Weyl}.

We note that (\ref{AIfix}) precludes both $W_2$ and $A_I$ from vanishing simultaneously for any $J\neq 0$, which would be necessary to have a regular solution at the origin 
as can be seen from Eqs.~\eqref{int_soln_a} and \eqref{int_soln_d}. Conversely, if we had demanded that the perturbation functions were regular at the origin (requiring $W_2=A_I=0$), 
then Eq.~(\ref{AIfix}) forces $J=0$, which is no rotation at all. 

The reason for this is that the $\ell \!=\!2$ quadrupole perturbations for $h_2$ (exterior)
or $rh_2$ (interior) satisfy an associated Legendre differential equation with argument $z= r/M -1$ or $z=1/Hr$ respectively, with a
particular inhomogeneous term depending upon $\w^{\prime\, 2}$. The singular points $z= (1, \infty)$ of the Legendre equation correspond to  $\!r=\!2M\!=\!R_{_S}$ and $r\!=\!\infty$ for the exterior solution, and to
$r\!=\!1/H\!=\!R_{_H}$ and $r\!=\!0$ for the interior solution. 
The homogeneous solutions $Q^2_2(z)$ or $P^2_2(z)$ are singular at $z\!=\!1$ or $z\!=\!\infty$, respectively, {\it cf.}~(\ref{Q22}).
 Requiring that the quadrupole perturbations remain finite at these singular points for both the exterior and interior solutions would set all four integration constants 
$A_{_E}\!=\!B_{_E}\!=\!A_{_I}\!=\!B_{_I}\!=\!0$ in (\ref{h2ext})-(\ref{v2ext}) and (\ref{int_soln_v})-(\ref{m2int}).
But then requiring the junction condition of continuity of the induced metric (\ref{surfmet})-(\ref{Sigdef}) 
at $r\!=\!R_{_H}\!=\!R_{_S}$ requires 
(\ref{eq:k2H_int_lim}) and (\ref{eq:k2H_ext_lim}) be equal and thus $W_2^2 = 2J^2$, which if nonzero gives a diverging
interior solution for $\w, h_2, k_2, m_2$ in (\ref{int_soln_a}) and (\ref{int_soln2}). Hence the only way to eliminate all divergences in the quadrupole functions is to set $W_2\!=\!J\!=\!0$, {\it i.e.}~no rotation at all.


Thus it is impossible to obtain an interior solution for a slowly rotating gravastar with finite values 
at the origin and the horizon, matched there to an asymptotically flat finite solution at infinity 
in the Hartle-Thorne framework---if the equation of state of the interior is unchanged from $p\!=\! -\r$ 
of the nonrotating case. We discuss the implications of this and how the divergences might be 
eliminated in a more complete treatment in the Discussion of Sec.~\ref{Sec:SumDisc} and Conclusions
of Sec.~\ref{Sec:Conclusions}.

\section{Surface Gravity and Stress Tensor of the Null Horizon Surface}
\label{Sec:SurfTens}

For timelike or spacelike hypersurfaces, one may relate the surface stress-energy to the discontinuity of the extrinsic curvature defined through the surface 
normal vector $\bn$ \cite{Israel:1966a}. For a null surface where $\bn \cdot \bn=0$, the extrinsic curvature vanishes, and the second junction conditions 
cannot be applied as formulated in \cite{Israel:1966a}. In the Barrab\`es-Israel method \cite{BarrabesIsrael:1991}, the surface stress-energy for a null surface is defined in terms of the 
discontinuity of an ``oblique" curvature defined in terms of a second transverse vector $\bN$. As discussed in Paper I  \cite{axihorizon}, this method 
also does not apply to the case at hand here where $j = \sqrt{h/f}$ and $\sqrt{-g}$ are discontinuous. Instead Paper I  \cite{axihorizon} contains a derivation of the second 
junction conditions for a null surface directly from the Einstein tensor density $\sqrt{-g} \, G^\m_{\ \,\n}$ for the horizon of an arbitrary metric of the form~\eqref{axisymstat}. Following Paper I,  the stress tensor localized on the null horizon surface is \cite{axihorizon}
\begin{align}
^{(\S )}T^a_{\ b}\, e^{\a + \n}  = \cS^a_{\ b} \, \d(r-\RH)
\label{SurfTS}
\end{align}
for $a,b=t,\th,\f$, where $^{(\S )}T^a_{\ b}$ is a well-defined Dirac $\d$-distribution with respect to the volume integration measure $\sqrt{-g}\, dt \,dr\, d\th \,d\f$,
analogously to (\ref{surfTsph}).

Under transformations of the hypersurface coordinates $(t,\th,\f)$, $e^{\b+\j} \cS^a_{\ b}$ transforms as a tensor density, while under transformations involving the surface coordinates $(\th,\f)$ only, $\cS^A_{\ B}$ is a tensor, $\cS^t_{\ t}$ is a scalar, and $\cS^t_{\ B}$, $\cS^A_{\ t}$ are vectors, for $A,B=\th,\f$. The null-surface stress tensor has nonzero components \cite{axihorizon}
\vspace{-4mm}
\begin{subequations}
\label{axisurf}
\begin{align}
8 \pi \cS^t_{\ t} &= -\w_H [\mathcal{J}]\\
8 \pi  \cS^t_{\ \f} &= [\mathcal{J}]\\
8 \pi  \cS^\f_{\ t} &= -\w_H[\k]-\w_H^2[\mathcal{J}]\label{Sphit}\\
8 \pi  \cS^\f_{\ \f} &=  [\k]+\w_H[\mathcal{J}] \\
8 \pi \cS^\th_{\ \th} &=   [\k]
\end{align}
\end{subequations}
where the surface gravity $\k$ is to second order in the rotation
\begin{align}
 \k & = \frac{1}{2} \, e^{-\a-\n}  \frac{\pa}{\pa r} \, e^{2\n} 
= \frac{1}{2j} \frac{dh}{dr} \left\{1 + h_0 + h_2 P_2 - \left (\frac{m_0 + m_2 P_2}{r-2m}\right)\right\} + \frac{h}{j} \left(h_0' + h_2' P_2\right)
\label{kaprot}
\end{align}
and $\cJ$, which is first order in the rotation, is
\begin{align}
\cJ =-\sdfrac{1}{2} \,e^{2\j} \, e^{-\a-\n}  \ \frac{\pa \w}{\pa r} = -\frac{j}{2}\, r^2 \sin^2\!\th\, \frac{d\w}{dr}
\label{calJ}
\end{align}
The square brackets in (\ref{axisurf}) denote the discontinuities in these quantities at the horizon, {\it i.e.} for a function $F(r,\th)$,
\vspace{-4mm}
\begin{align}
[F] = F_{+} - F_{-} 
\end{align}
with
\vspace{-3mm}
\begin{align}
F_{\pm} \equiv \lim_{r\to \RH^{\pm}} F(r,\th) \,.
\end{align}

With the interior and exterior perturbations given by the results of the previous two sections, namely the finiteness conditions $B_E=B_I=0$, and the conditions \eqref{eq:dRH_ext}, \eqref{eq:elim_E0}, \eqref{AIfix}, \eqref{W1fix} on the horizon,
the surface gravity as the horizon boundary at $r=\RH=\RS$ is approached from the interior and exterior is, to second order in the rotation
\begin{subequations}
\begin{align}
\k_- &= \lim_{r\to \RH^-} \k (r) = -\frac{1}{j\RH} \,\left(1 + C - \frac{3 \,\d M_I}{\RH}  + \frac{3 j^2 W_2^2}{2\RH^4}\right)  \\
\k_+ &= \lim_{r\to \RH^+} \k (r) = \frac{1}{2 \RH} \,\left( 1 - \frac{6J^2}{\RH^4}\right) \,.
\label{eq:kappa_plus}
\end{align}
\end{subequations}
The discontinuity of the surface gravity at the horizon is therefore
\begin{align}
[\k] = \k_+ - \k_- = \left(\frac{1}{2} + \frac{1+C}{j} \right) \frac{1}{\RH}  - \frac{3J^2}{\RH^5}- \frac{3\, \d M_I}{j \RH^2}+ \frac{3jW_2^2}{2\RH^5}
\label{kdisc_j}
\end{align}
which for a gravastar ($j=2$) becomes
\begin{align}
[\k] = \frac{1}{\RH} + \frac{C}{2 \RH} - \frac{3\, \d M_I}{2 \RH^2} - \frac{3J^2-3W_2^2}{\RH^5}\,.
\label{kdisc}
\end{align}
Similarly $\cJ$ approached from the interior and exterior is
\begin{subequations}
\begin{align}
\cJ_- &= \lim_{r\to \RH^-} \cJ (r) =  \frac{3jW_2}{2\RH^2\!}\,\sin^2\!\th\\
\cJ_+ &= \lim_{r\to \RH^+} \cJ (r) = \frac{3J}{\RH^2\!}\, \sin^2\!\th
\end{align}
\end{subequations}
so that its discontinuity at the horizon is
\begin{align}
[\cJ] =  \cJ_+ - \cJ_- = \frac{3(2J-jW_2)}{2\RH^2}\, \sin ^2\!\th
 = \frac{3(J-W_2)}{\RH^2}\, \sin ^2\!\th
\label{discJ}
\end{align}
with the latter expression for a gravastar ($j=2$). 

Dropping the second $\cO(J^3)$ term in (\ref{Sphit}),  Eqs.~(\ref{axisurf}) then give for a gravastar, 
\begin{subequations}
\label{surfstress}
\begin{align}
8 \pi  \cS^t_{\ t} &=-\frac{6 J(J-W_2)}{\RH^5}\,  \sin^2\!\th \\
8 \pi  \cS^t_{\ \f} &=\frac{3(J-W_2)}{\RH^2}\,  \sin^2\!\th \, \\
8 \pi  \cS^\f_{\ t} &=-\frac{2 J}{\, \RH^4} \\
8 \pi  \cS^\f_{\ \f} &=\frac{1}{\RH} +  \frac{C}{2 \RH} - \frac{3J^2}{\RH^5} - \frac{3 \, \d M_I}{2\RH^2} + \frac{3W_2^2}{\RH^5}+\frac{6 J(J-W_2)}{\RH^5}\,  \sin^2\!\th  \\
8 \pi  \cS^\th_{\ \th} &= \frac{1}{\RH} + \frac{C}{2 \RH}  - \frac{3 J^2}{\RH^5} - \frac{3 \, \d M_I}{2\RH^2} + \frac{3W_2^2}{\RH^5}
\end{align}
\end{subequations}
in terms of the integration constants $J, C, W_2, \d M_I$,  after using the conditions \eqref{eq:dRH_ext}, \eqref{eq:elim_E0}, \eqref{AIfix}, \eqref{W1fix} on the horizon at $r=\RH=\RS$. In the nonrotating limit, with $J,  C, W_2$, and $\d M_I$ set to zero, the surface stress-energy tensor of the nonrotating gravastar is recovered \cite{MazurMottola:2015}. 
 
The surface stress components in \eqref{surfstress} indicate that for general values of the constants $W_2$, $\d M_I$, and $C$, there are nonvanishing anisotropic 
surface stresses, an azimuthal current, and an energy density localized on the horizon surface. The $\cS^\f_{\ \f}$ and $\cS^\th_{\ \th}$ terms are the azimuthal 
and latitudinal stresses, containing a zeroth order (nonrotating) term with second order corrections. Because the zeroth order term is associated with the spherically 
symmetric unperturbed system, it is the same in the $\th$ and $\f$ components, but the second order corrections may differ (equal only if $W_2=J$). The azimuthal current terms 
$\cS^t_{\ \f}$ and $\cS^\f_{\ t}$ are first order in the rotation. $S^\f_{\ t}$ is nonzero for a rotating gravastar ($J\ne0$), while $S^t_{\ \f}$ is nonzero in general and 
vanishing only if  $W_2=J$. The energy density term $\cS^t_{\ t}$ is second order in the rotation, and vanishes if $W_2=J$.
 
\section{Gravastar Mass, Angular Momentum, and Moment of Inertia}
\label{Sec:Komar}

Section V of Paper I \cite{axihorizon} gives the integrals for the Komar mass and angular momentum functions for an arbitrary metric of type (\ref{axisymstat}).
The Komar functions at some radius are given by a surface integral at that radius, which may be decomposed by Stokes' theorem into the contribution from a surface 
of smaller radius and a volume integral of appropriate energy-momentum components. 

\subsection{Komar mass}

The Komar mass function is \cite{axihorizon} 
\begin{subequations}
\label{KomarMass}
\begin{align}
M_K(r) &=  \frac{1}{4 \pi }\int_{\pa V_+}\!  (\k + \w \mathcal{J})\, dA \label{KomarMass1} \\
& =\int_V \sqrt{-g}\  \Big(\! \!-T^t_{\ t} + T^r_{\ r}  + T^\th_{\ \th}  + T^\f_{\ \f}  \Big)\, dr\,d\th\,d\f\, +  \frac{1}{4 \pi }\int_{\pa V_-}\!  (\k + \w \mathcal{J})\, dA
\label{KomarMass2}
\end{align}
\end{subequations}
where $V$ is the three-volume at fixed $t$ enclosed by two-surfaces $\pa V_-$ and $\pa V_+$ at $r_-$ and $r_+$ in coordinates  (\ref{axisymstat}).

In the exterior the Komar mass is easily found to be
\begin{align}
\MKext =M  + \d M = \MM\,,\qquad r> \RH
\label{Massex}
\end{align}
which is the constant total mass of the exterior Kerr solution, as expected. Since the exterior is a vacuum solution with $T^{\m}_{\ \n} =0$, 
there is no volume contribution in~\eqref{KomarMass2} and no $r$ dependence to the Komar mass function in the exterior. 
Since the exterior is Kerr, we may compare Eq.~(\ref{Massex}) to the Christodoulou formula for a Kerr BH \cite{Christo:1970}
\vspace{-3mm}
\begin{align}
    \MKext=\sqrt{M_{irr}^2+\frac{J^2}{4M_{irr}^2}}\approx M_{irr}+\frac{J^2}{8M_{irr}^3}
    \label{KerrMM}
\end{align}
where $M_{irr}$ is the irreducible mass and the final equality comes from expanding for small $J$ at constant $M_{irr}$. With $h_0$ finite and $\d M$ given by (\ref{eq:dRH_ext}), one can see from Eqs.~(\ref{Massex}) and (\ref{KerrMM}) that the unperturbed mass of the nonrotating gravastar $M$ equals the irreducible mass of the Kerr BH, as implied by the $\RH=\RS$ condition and the definition of the irreducible mass $4\pi \RH^2 = 16 \pi M_{irr}^2$.

The two-surface integrals in (\ref{KomarMass}) may be evaluated at fixed $r$ with the induced area element
\begin{align}
dA = e^{\b + \j} d\th\,d\f = r^2\,\sin\th\, \Big[1 + 2k_2(r) P_2(\cos\th)\Big] d\th\,d\f 
\end{align}
and the total area of the horizon is
\begin{align}
A = \int e^{\b + \j} d\th\,d\f =  \int d\th\,d\f \, R_{_H}^2 \sin\th \Big[1 + 2 k_2(R_{_H}) P_2 (\cos\th) \Big] = 4\pi R_{_H}^2= 16\pi M_{irr}^2=16\pi M^2
\label{Area}
\end{align}
since the quadrupole term does not contribute. The final equality of (\ref{Area}) makes use of the condition 
of finite $h_0$ and matching at $\RH=\RS$.

Substituting the perturbations $h_0, m_0, h_2,m_2$ for the interior solution (\ref{interiorsoln}), and performing the surface integral $\pa V_+$ at $r$,
the quadrupole terms proportional to $P_2(\cos\th)$ again do not contribute, and we obtain
\begin{align}
\frac{1}{4 \pi }\int_{\pa V_+}\!  \k dA &= -\frac{r^3}{j\RH^2} \bigg(1+C+ \frac{ 8 \pi \EE_0\RH^2}{3} \bigg) - \frac{jW_2^2}{r^3}  + \frac{\d M_I}{j}
\label{Masskint}
\end{align}
retaining terms up to second order in the rotation. Likewise for the $\w\cJ$ term in (\ref{KomarMass1}), we obtain
\begin{align}
\frac{1}{4 \pi }\int_{\pa V_+}\!  \w \cJ \, dA = j W_2 \left(W_1 + \frac{W_2}{r^3} \right) 
\end{align}
correct to second order. Combining this with (\ref{Masskint}) we find the expression for the Komar mass in the interior, 
\begin{align}
\MKint(r)=  -\frac{r^3}{j\RH^2} \bigg(1+C+ \frac{ 8 \pi \EE_0\RH^2}{3}\bigg) + \frac{j^2W_1W_2 + \d M_I}{j}
 \,,\qquad 0<r< \RH\,.
\label{MassKInt}
\end{align}
The $r$ dependent term in~\eqref{MassKInt} can be identified as the volume contribution from (\ref{KomarMass2}), 
\begin{align}
& \int_V \sqrt{-g}\  \Big(\! \!-T^t_{\ t} + T^r_{\ r}  + T^\th_{\ \th}  + T^\f_{\ \f}  \Big)\, dr\,d\th\,d\f  \nn
& = - 2(\r + \EE_0) \int_0^r dr \int d\th \, d\f  \frac{r^2\sin\th}{j}\left( 1 + h_0 + h_2P_2 + \frac{m_0 +m_2 P_2}{r (1-H^2r^2)} + 2 k_2 P_2\right)\nn
&=  -\frac{r^3}{j\RH^2} \bigg(1+C+ \frac{ 8 \pi \EE_0\RH^2}{3}\bigg)\,.
\end{align}
Extending the latter integration over the whole interior volume $0<r<\RH$ gives the total interior volume contribution to the Komar mass
\begin{align}
\MKvolume & 
=  -\frac{\RH}{j} \bigg(1+C - \frac{2\,\d M_I}{\RH} +\frac{j^2W_2^2}{2\RH^4}  \bigg) 
\label{volintM}
\end{align}
where $\RH=\RS$ and $\EE_0$ from (\ref{eq:elim_E0}) has been used.

The $r$ independent term in~\eqref{MassKInt}, {\it i.e.},
\begin{align}
\MKorigin = \frac{j^2W_1W_2 + \d M_I}{j}
\label{mk0}
\end{align}
which is present at arbitrarily small $r$, can be ascribed to $\d$-function contribution to the volume integral in Eq.~(\ref{KomarMass2}) at the origin
by 
\begin{align}
\frac{1}{4\pi}\int d\th d\phi \sqrt{-g}\sq K^{t}_{(t)}\Big\vert_{\rm origin} = \int d\th d\phi \sqrt{-g}  \Big(\! \!-T^t_{\ t} + T^r_{\ r}  + T^\th_{\ \th}  + T^\f_{\ \f}  \Big)\Big\vert_{\rm origin}  =\MKorigin \d (r) \, \label{pointmass}
\end{align}
where $K^{t}_{(t)}$ is the $t$ component of the Killing vector of time translation $\pa/\pa t$.
Thus the general solution to the slow rotation perturbation equations allows for the constant contribution $\MKorigin$ to the Komar mass in (\ref{mk0}), much as integration of the mass function 
equation $m' = 4 \pi r^2 \,\r$  allows for an integration constant and mass at the origin. 
Requiring the absence of any such $\d$-function at the origin requires
\begin{align}
\d M_I = - j^2 W_1 W_2
\label{W2E0}
\end{align}
fixing one of the remaining three constants in the interior gravastar solution.

The contribution to the Komar mass of the arbitrarily thin surface layer stress energy localized on the horizon is
\begin{align}
\MKsurface & = \int_{0}^{\pi}\int_0^{2\pi}\int_{\RH-\epsilon}^{\RH+\epsilon} d\theta d\phi dr \sqrt{-g} \left(-T^t_{\ \,t} + T^\th_{\ \,\th} + T^\f_{\ \,\f} \right)
 \nn & = 2\pi \int_{0}^{\pi}\int_{\RH-\epsilon}^{\RH+\epsilon} d\theta dr\, e^{\j + \b}\, \d (r-\RH) \left(-\cS^t_{\ t} + \cS^\th_{\ \th} + \cS^\f_{\ \f} \right)\nn
&=\frac{1}{4\pi} \int \Big( \big[\kappa\big] + \w_H \big[\cJ\big]\Big) \, dA  
= \MKext- \MKvolume - \MKorigin
\nn
& = \left( \frac{1}{2}+\frac{1+C}{j} \right) \RH + \frac{J^2}{\RH^3} + \frac{jW_2^2}{2\RH^3} - \frac{3\, \d M_I+j^2W_1W_2}{j} 
\label{Msurf}
\end{align}
where $\RH=\RS$ and $\d M,\EE_0$ from (\ref{eq:dRH_ext}) and (\ref{eq:elim_E0}) have been used.

The total mass in the exterior
\begin{align}
\MKext = \MKorigin +\MKvolume + \MKsurface= \MM
\label{Mtotal}
\end{align}
is the sum of the contributions from the interior volume, the possible $\d$-function at the origin, and the surface layer, independently of the conditions 
(\ref{W2E0}). If (\ref{W2E0}) is satisfied, the contribution from the origin is absent: $\MKorigin=0$. Note also that changing the integration constant 
$C$ changes the contributions to the interior volume and surface terms separately, but drops out of the sum (\ref{Mtotal}). 

\subsection{Komar angular momentum}

The Komar angular momentum is expressed similarly to the Komar mass as \cite{axihorizon}
\begin{align}
J_K(r) =  \frac{1}{8 \pi } \int_{\pa V_+}\!  \! \mathcal{J} \, dA = \int_V  \sqrt{-g}\ T^t_{\ \f} \,dr\,d\th\,d\f +\frac{1}{8 \pi } \int_{\pa V_-}\!  \!\mathcal{J} \, dA 
\label{KomarAngmom}
\end{align}
which shows that $\cJ/8\pi$ carries the interpretation of angular momentum per unit surface area. Using (\ref{calJ}), the surface integral for the interior solution is
\begin{align}
\JKint = \frac{1}{8\pi } \int_0^{2\pi} d\f\int_0^\pi d\th  \,  r^2 \sin^3\!\th \left(\frac{3jW_2}{2r^2}\right) = \frac{jW_2}{2}
\end{align}
which is first order in the rotation. Since this is independent of $r$, and $T^t_{\ \,\f} = 0$ in the interior, the volume integral term for angular momentum in~(\ref{KomarAngmom}) gives no contribution,
\begin{align}
\JKvolume = 0
\label{eq:Jvolume}
\end{align}
but the constant contribution
\vspace{-4mm}
\begin{align}
\JKorigin= \frac{jW_2}{2}
\label{eq:Jorigin}
\end{align}
can be ascribed to the volume term from Eq.~(\ref{KomarAngmom}) as a $\d$-function contribution at the origin,
\begin{align}
-\frac{1}{4\pi}\int d\th d\phi\sqrt{-g}\sq K^t_{(\f)}\Big\vert_{\rm origin} \, = 2 \int d\th d\phi \sqrt{-g} T^t_{\ \,\f}\Big\vert_{\rm origin} =2\, \JKorigin  \d (r) 
\end{align}
where $K^{t}_{(\f)}$ is the $t$ component of the Killing vector of azimuthal symmetry $\pa/\pa \f$.
Eliminating this $\d$-function contribution to the angular momentum requires $W_2 = 0$. Eliminating both the Komar angular momentum and Komar mass delta functions at the origin thus requires
\begin{equation}
    \d M_I = W_2=0
    \label{nodeltas}
\end{equation}
so that the only remaining integration constant of the interior solution is $C$.

The Komar angular momentum function in the exterior is
\begin{align}
\JKext&= - \frac{1}{8} \int_0^\pi d\th  \sqrt{\frac{h}{f}}\, r^4 \sin^3\!\th \,\frac{d}{d r}\!\left(\frac{2J}{r^3}\right)= J
\end{align}
which is the expected constant total angular momentum of the Kerr exterior solution. Finally, we have the angular momentum contribution of the surface layer
\begin{align}
&\JKsurface =\int_{0}^{\pi}\int_0^{2\pi}\int_{\RH-\e}^{\RH+\e} d\th d\phi dr \sqrt{-g}\, T^t_{\ \f} = 
2\pi \int_{0}^{\pi}\int_{\RH-\e}^{\RH+\e} d\th dr\,  e^{\j + \b}\, \d (r-\RH)\, \cS^t_{\ \f} \nn
&\hspace{5mm} =\frac{1}{8\pi} \int d\phi d\th  e^{\j + \b}\,\big[ \cJ \big]  = \frac{3(2J-jW_2)}{8\RH^2} \int_{0}^{\pi}\int_{\RH-\e}^{\RH+\e} d\th dr\, r^2 \sin^3 \th\, \d (r-\RH)\, =J-\frac{jW_2}{2}
\label{Jsurf}
\end{align}
to this order in rotation. This shows that the total Komar angular momentum of the gravastar is the sum of contributions from the surface and the origin,
\begin{align}
\JKext=\JKorigin + \JKsurface = J
\end{align}
If $W_2=0$, then $\JKorigin=0$ and the entire angular momentum 
of the slowly rotating gravastar comes from the rotating surface layer localized on the horizon surface.

\subsection{Moment of inertia}

The moment of inertia $I$ of the rotating gravastar is
\begin{align}
I = \frac{J}{\w_H} = 4 G^2 M^3 = M \RH^2
\label{Inertia}
\end{align}
which coincides with that of a Kerr BH. This might have been expected from the fact that the external geometry is identical to the Kerr BH geometry. With $W_2=0$,
the interior de Sitter condensate carries no angular momentum from (\ref{eq:Jvolume}) and~\eqref{eq:Jorigin}, and all the angular momentum is
carried by the rotating horizon surface from (\ref{Jsurf}). Unlike a BH where there is no mass-energy at all at the
horizon to rotate or to give rise to such a moment of inertia, the rotating gravastar has a well-defined stress energy (\ref{surfstress}) on its rotating surface where all the angular momentum resides, and it is this rotating mass-energy that gives rise to (\ref{Inertia}) by straightforward evaluation of
the relevant surface integral (\ref{Jsurf}). 

\section{Summary and Discussion}
\label{Sec:SumDisc}

With the $\d$-functions contributions to the Komar mass and angular momentum
at the origin set to zero ($\MKorigin=0$ and $\JKorigin=0$), and the condition that $h_0$ is finite at the gravastar surface, all interior integration constants for the slowly rotating gravastar are fixed by Eqs.~\eqref{eq:dRH_ext}, \eqref{eq:elim_E0}, \eqref{AIfix}, \eqref{W1fix}, \eqref{nodeltas}, except for $C$. The interior slowly rotating gravastar solution in this case simplifies considerably,
becoming
\vspace{-4mm}
\begin{subequations}
	\label{interiorsolnfin}
	\begin{align}
	\w&=\frac{2J}{\RH^3}= \w_H \label{omfin}\\
	m_0&= 0 \label{m0fin}\\
	h_0&=C\label{h0fin}\\
	m_2&=-\frac{2J^2(r^2-\RH^2)^2}{r^2\RH^5}\label{m2fin}\\
	h_2&= \frac{2J^2}{r\RH} \left(\frac{1}{r^2} - \frac{1}{\RH^2}\right)\label{h2fin}\\
	k_2&=-\frac{2J^2}{r\RH}\left(\frac{1}{r^2} + \frac{1}{\RH^2}\right)\label{k2fin}
	\end{align}
\end{subequations}
with the exterior solution given by Eqs.~(\ref{extfinal}).  The surface stress tensor is obtained by setting $\d M_I=W_2=0$ in Eqs.~(\ref{surfstress}), {\it i.e.}
\vspace{-4mm}
\begin{subequations}
	\label{surfstressfin}
	\begin{align}
	8 \pi  \cS^t_{\ t} &=-\frac{6 J^2}{\RH^5}\,  \sin^2\!\th \\
	8 \pi  \cS^t_{\ \f} &=\frac{3J}{\RH^2}\,  \sin^2\!\th \, \\
	8 \pi  \cS^\f_{\ t} &=-\frac{2 J}{\, \RH^4} \\
	8 \pi  \cS^\f_{\ \f} &=\frac{1}{\RH} +  \frac{C}{2 \RH} - \frac{3J^2}{\RH^5} +\frac{6 J^2}{\RH^5}\,  \sin^2\!\th  \\
	8 \pi  \cS^\th_{\ \th} &= \frac{1}{\RH} + \frac{C}{2 \RH}  - \frac{3 J^2}{\RH^5}\,.
	\end{align}
\end{subequations}

The constant $C$ does not appear in the bulk ($r < \RH$) stress-energy tensor or curvature tensor of the interior solution. Referring to the metric 
line element (\ref{kingmet}), we observe that the constant $1+ C$ is a rescaling of the time coordinate of the interior, and thus appears to amount 
to a choice of coordinates. It is indeed similar to the constant $1+ C_{_E}$ of (\ref{h0Cext}), which is a rescaling of the time coordinate of the exterior 
solution that is fixed by the condition that the asymptotically flat line element assumes the standard Minkowski form. It is also similar
to the $j$ factor, which rescales the interior $g_{tt}$ component of the metric in the nonrotating gravastar. In that spherically symmetric 
case the value $j\!=\!2$ is fixed by taking the horizon limit $R \to \RH$ of the Schwarzschild star~\cite{MazurMottola:2015}. Similarly, if the interior 
and exterior were continuously connected in an appropriate horizon limit, or if $f(r)$ were to be kept slightly different from zero near $r = \RH$ by a more detailed theory of the surface layer, we may expect that $C$ would be fixed by the one condition on the Minkowski static time coordinate as $r\to \infty$. 

As it is, in our present treatment because of the lightlike null surface intervening where $f(\RH) = 0$, the time coordinate of the interior solution 
becomes disconnected from the exterior, and $C$ remains a free parameter, which appears in the interior bulk Komar mass (\ref{MassKInt}), the surface 
stress tensor (\ref{surfstress}) and surface contribution (\ref{Msurf}) separately, but drops out in their sum $\MM$ for the total mass (\ref{Mtotal}).

It is important to note that our results are contingent on the surface layer being located on the null horizon and infinitesimally thin, following the matching method presented in Paper I~\cite{axihorizon}.  
An infinitesimal surface which is located away from the null horizon leads to different results and has been studied by other authors, to which we may explicitly compare our results. Previous works were based on matching the interior Hartle-Thorne perturbed de Sitter
to exterior geometries on a timelike \cite{Pani:2015,Uchikata:2015,UchiPani:2016} or spacelike \cite{UchiYoshida:2014} hypersurface $r\!=\!R\!=\!{\rm const}\neq 2M$. These authors made a different choice of constants for both the exterior and interior perturbations. They use conventional time scaling in the interior $j=1$ and do not consider $\EE_0$, but allow $B_{_E}, B_I$ to be nonzero, since they did not require the
perturbations to be finite at the horizon. They set $A_I = W_2 = \d M_I =0$ in order to have finiteness of the perturbations at the origin $r=0$.
Comparing our solution in (\ref{int_soln}) with Eqs.~(2.15)-(2.19) of Ref.~\cite{UchiPani:2016}, we see that the  interior solution of Ref.~\cite{UchiPani:2016} amounts to 
choosing $W_2 \!=\!\d M_{I} \!=\! A_I \!=\! \EE_0 \!=\! 0$, with arbitrary $W_1, B_I, C$ in the present notation. With these choices, $k_2(r)$ diverges as $r_{_S} \to 1/H$ unless it vanishes identically, {\it i.e.} also $B_I=0$.
The latter option would then set $m_0, h_2, k_2, m_2$ all to zero identically, and leave only $\w=W_1, h_0=C$ as constants in the interior. But this interior cannot be matched to the uniquely well behaved Kerr exterior on the horizon. 

In contrast, finiteness and matching on the null horizon 
surface requires (\ref{BIfix}) and leaves a generally nontrivial interior solution for the perturbations, which however may still diverge at the origin if either $A_I$ or $W_2$
are nonvanishing. The junction conditions of matching on the horizon fix the relation (\ref{AIfix}), which precludes setting both $A_I$ and $W_2$ to zero if $J \neq 0$.

The earlier authors \cite{UchiYoshida:2014,Pani:2015,Uchikata:2015,UchiPani:2016} applied the Israel junction conditions, which do not apply in the limit $R \to 2M$, as they produce either a vanishing
or a divergent result depending on whether the extrinsic curvature is computed with both lower contravariant indices, or with one
of its indices raised, due to the fact that the induced metric on the null horizon is degenerate. The relation between
the surface stress tensors of these previous approaches and the present one is given by (\ref{Isrmod}) \cite{axihorizon}.

Another paper \cite{Rutkowski:2009} uses a different perturbation method in Eddington-Finklestein coordinates on rotating gravastars derived directly from the Schwarzchild star limit. This work finds regular perturbative solutions to second order, which fail only at third order. However, the analysis differs from ours in that the surface of the gravastar is allowed to shift away from being null in the rotating case, whereas we have performed the matching on a null surface in both the static and rotating cases. 

Our analysis also differs from \cite{Posada:2016}, where a sub-Buchdahl Schwarzschild star ($R < 9M/4$) was studied numerically in the Hartle formalism by matching the interior and exterior metrics continuously on a timelike hypersurface and approaching the horizon limit numerically. There it was suggested that the perturbative corrections were everywhere finite in the gravastar limit. However, below the Buchdahl bound the equation for frame dragging solved numerically, Eq.~(50) of~\cite{Posada:2016}, does not follow  from the change of variables given in the definitions in Eqs.~(45)-(49) and the original frame-dragging equation in $r$ coordinates, {\it cf.}~Eq.~(25). Hence the results in~\cite{Posada:2016} would have to be recomputed before comparing them to our work, although the same result (\ref{Inertia}) for the moment of inertia was obtained.

A major result of our analysis is that although one may remove the $\d$-function contributions at the origin to both the Komar mass function 
$\MKorigin$ and angular momentum $\JKorigin$ by setting  $\d M_I = W_2= 0$, the quadrupole perturbations $m_2$, $h_2$, and $k_2$ 
remain quadratically or cubically divergent at the origin for $J \neq 0$. The perturbations $m_2$ and $k_2$ in \eqref{eq:int_m2} and~\eqref{eq:int_k2} become of order unity at 
\vspace{-4mm}
\begin{align}
r_{\rm min} \sim\left(\frac{2J^2}{\RH}\right)^{\!\frac{1}{3}} 
\end{align}
and hence the perturbative expansion breaks down for $r \lesssim r_{\rm min}$ (or possibly $r \lesssim \sqrt{|W_2|}$ if one allows for
nonzero $W_2$). There are two possible reasons for this breakdown and corresponding possibilities for removing the divergences at $r=0$.

The first possibility is that the perturbative expansion in slow rotation $J= a M$ is nonuniform in $r$, and invalid near the origin.
If a diverging $a^2/r^2$ behavior is replaced by any smooth function $F(r/a)$ in a more complete solution,
where $F$ tends to unity for large values of its argument, but to zero rapidly enough as $r \to 0$, the divergence
at the origin would be removed. For example
\begin{align}
F\left(\sdfrac{r}{a}\right) = \frac{a^2 r^{2n -2}}{(a^2 + r^2)^n}
\label{Feg}
\end{align}
for $n > 1$ is finite both for $a=0$, and as $r\to 0$ for any $a >0$, where it behaves like $(r/a)^{2n-2}\to 0$
near the origin. However, expanding this function to order $a^2$ yields a leading term, $a^2/r^2$ which diverges as $r\to 0$. Thus
the diverging behavior at $r=0$ in the 
slow rotation expansion may imply that the true solution for the interior metric of a rotating gravastar, though everywhere
regular, does not admit a uniform regular expansion in the rotation parameter near the origin.  We have not found an exact rotating 
metric in the literature with $p=-\r$ equation of state and regular metric functions which expands into our solution 
in precisely this way, although there are solutions with some similar properties of $p=-\r$ equation of state and diverging expansion, 
such as the modified Carter solution presented in~\cite{ChaplineMarecki:2008}.

Another method which is been used in the literature to generate possible interior solutions to the Kerr BH 
(see {\it e.g.} \cite{Burinskii2002})  and manifests divergences on expansion is the Gurses-Gursey generalization \cite{Gurses:1975vu} of the Newman-Janis algorithm \cite{Newman:1965tw}. The Gurses-Gursey rotating system has the 
line element
\begin{align}
\hspace{-3mm}ds^2=-\left(1-\sdfrac{2 r m}{\S}\right)dt^2+\sdfrac{\S }{\D }dr^2 + \S\, d\th^2+\sin^2\!\th \left(\sdfrac{2 a^2 rm \sin^2\!\th}{\S }+a^2+r^2\!\right)d\f^2 -\sdfrac{4 a  r m\sin^2\!\th\!}{\S }\ dtd\f
    \label{spinmet}
\end{align}
where $\S(r,\th)=r^2+a^2\cos^2\!\th$, $\D(r)=r^2+a^2-2r m(r)$, and $m(r)$ is the enclosed mass in the spherically 
symmetric system, which may be taken to be $H^2r^3/2$ for the nonrotating de Sitter gravastar interior. Although this
metric is completely regular at $r=0$, expanding the $1/\S$ and $1/\D$ functions to order $a^2$ will generate spurious
divergences, similar to (\ref{Feg}). However, this rotating geometry \cite{Ibohal:2004kk,2006PhLB..639..368D, Azreg-Ainou:2014nra,deUrreta:2015nla}  is a solution of Einstein's equations with an energy-momentum
tensor that does {\it not} satisfy the $p=-\r$ dark energy equation of state \cite{2021arXiv210402255B}. 

This brings us to the second possibility for removing the divergence at the origin, which is that the preservation of the $p=-\r$ equation of state
may be too restrictive for an ultra-compact rotating gravastar. Our analysis is based
on the assumption, eminently reasonable for most fluids, that the local thermodynamic equilibrium equation of state would not be modified if the
fluid sphere is set into slow rotation. However the dark energy equation of state, at the extreme limit of satisfying the weak energy
condition $\r  + p \ge 0$, is not that of an ordinary fluid, and need not follow the expectation of local thermodynamic equilibrium
under small perturbations familiar from classical fluids. If the equation of state is altered from dark energy form, as for instance in the Gurses-Gursey system Eq.~(\ref{spinmet}), then there may be angular momentum density distributed within the interior rather than purely on the surface as in our solution. Indeed if the interior of the gravastar is a gravitational condensate as 
originally proposed in  \cite{MazurMottola:2001,MazurMottola:2004}, one might expect the development of vortex filaments first
at the origin, and then a finite vortex density at higher values of $cJ/GM^2$. Such vortices would alter the equation of state from 
its pure vacuum dark energy condensate form, altering the solution markedly, deep in the interior. One might speculate that 
even in this case the physical properties of the surface layer may not be radically changed at the horizon from that given in the 
present work, at least to second order in the rotation parameter $cJ/GM^2$.

\section{Conclusions}
\label{Sec:Conclusions}

In this paper we have derived and solved analytically the Einstein equations for a slowly rotating gravastar by expanding
the spherically symmetric gravastar solution of \cite{MazurMottola:2015} to second order in the angular momentum, 
following the methods of Refs.~\cite{Hartle:1967he, Hartle:1968si,Chandrasekhar:1974}, and assuming that the equation of state of the slowly rotating system is unchanged from that of the nonrotating solution.

In solving the perturbation equations for slow rotation, we applied the lightlike shell junction conditions of Paper I \cite{axihorizon} to the concrete case of a rotating gravastar. Previous studies in the Hartle-Thorne formalism have 
focused on the case of matching on a timelike or spacelike boundary~\cite{Pani:2015,Uchikata:2015,UchiPani:2016,UchiYoshida:2014,Reina:2015}.

The surface stresses and Komar integrals work as expected, with the modifications appropriate for matching on a null hypersurface. The $\d$-function distribution stress tensor on the horizon surface (\ref{SurfTS})-(\ref{axisurf})
accounts for the difference between the exterior and interior Komar mass and angular momentum functions. 

In the Komar description, the $p=-\r$ equation of state cannot support any angular momentum density and the angular momentum must 
either be localized entirely on the horizon or in singular concentration at the origin. With the conditions (\ref{nodeltas}) removing any $\d$-function contribution at the origin, the entire angular momentum $J$ of the rotating solution is carried by the physical surface of the gravastar at the horizon.

 Matching solutions on the horizon, we determine the interior metric and the surface stress tensor corrections of first and second order in the rotation.
The exterior solution for the perturbations (\ref{extfinal}) yields exactly the same geometry as that of a Kerr BH, expanded to second order in $a=J/M$, but expressed in Hartle-Thorne coordinates (\ref{kingmet}), which are related to 
more familiar Boyer-Lindquist coordinates by (\ref{BL2HT}). Thus an important conclusion of this work is

\vspace{-2mm}
{\setstretch{1.4}
\begin{itemize} [leftmargin=6mm]
\item Matching the interior of a slowly rotating gravastar to an asymptotically flat exterior vacuum solution of Einstein's equations at their horizons leads to an external geometry {\it identical} to that of a slowly rotating Kerr BH, notwithstanding a very different interior and the presence of nonzero energy density and stresses (\ref{surfstressfin}) localized on the physical surface, that takes the place of the Kerr BH horizon.
\end{itemize}}
\vspace{-3mm}
\noindent
Since the exterior of the slowly rotating gravastar is precisely that of the Kerr BH to order $a^2$, there is no possibility of detecting the difference to this order by any experiment or observation restricted to the large scale external geometry such as by accretion flows or light ring images.
Like a classical Kerr BH, a slowly rotating gravastar with a null surface stress tensor also has ``no hair" classically, up to possibly Planck scale corrections very near to the horizon surface. The fact that this surface remains at the horizon, and the moment of inertia (\ref{Inertia}) remains identical to that of a Kerr BH 
to this order also implies that a rotating gravastar defined this way will also have the same zero tidal deformability or Love number as a Kerr BH, at least to order $a^2$.

If the $\d$-function surface stress-energies were to be replaced by a microscopically thin shell of radial extent $\D r$, several studies~\cite{CarFraMasPaniRap:2017,ChirPosGued:2020} indicate that the corrections to this null result may be of order $[\log(\D r/M)]^{-1}$, so that even a $\D r$ of order the Planck length could 
potentially produce dimensionless Love numbers of a few tenths of a percent, which could lead to observable tidal effects that can distinguish ultra-compact objects from BHs in future gravitational wave detectors~\cite{UchiPani:2016,CarFraMasPaniRap:2017}. This question remains open for the universal null shell gravastars considered in this paper.

Assuming that the interior equation of state for a slowly rotating gravastar is unchanged from its nonrotating (\ref{EMT})-(\ref{eq:E0_E2}) form,
the regular matching of the exterior Kerr and the interior spacetime at the horizon gives Eq.~(\ref{AIfix}), which in our calculation implies a  
divergence of some perturbation functions at the origin. As explained in Sec.~\ref{Sec:Joining}, since the horizons of the interior and exterior geometries are at singular points of the differential equation governing the quadrupolar 
perturbations, the four conditions:
\vspace{-3mm}
\begin{enumerate}[label={(\roman*)}]
\itemsep-2.5mm
\item Finiteness at the origin,
\item Finiteness on the horizon,
\item Finiteness at asymptotic infinity ({\it i.e.} asymptotic flatness), and
\item Matching (first junction condition) on the horizon,
\end{enumerate}
\vspace{-3mm}
cannot be simultaneously satisfied if the equation of state in the interior is fixed to be $p=-\r$.
Thus a second important conclusion of this work is:
\vspace{-2mm}
{\setstretch{1.4}
\begin{itemize} [leftmargin=6mm]
\item There is no solution of the Hartle perturbation equations for a slowly rotating
gravastar that is regular both at the horizon and the origin, and can be matched on the horizon to  
an asymptotically flat exterior,  if the equation of state in the interior is fixed to be $p=-\r$  of the nonrotating gravastar.
\end{itemize}}
\vspace{-2mm}
\noindent
The divergences may indicate that the slow rotation expansion is nonuniform and breaks down as $r\to 0$, and/or that the assumption that the equation of state is unchanged from $p = -\r$ is overly restrictive for ultracompact objects such as gravastars,
and should be relaxed. If instead, the equation of state differs from the dark energy one anywhere in the interior, then the perturbation functions inside the gravastar
will also be different and so their matching to the exterior functions at the null surface may become possible, while still remaining finite
at the origin.

The corollary is that the slowly rotating gravastar model generated in this way by the Hartle-Thorne 
formalism~\cite{Hartle:1968si,Chandrasekhar:1974} from the spherically symmetric case is necessarily incomplete, and not fully satisfactory. The present work
represents an initial exploration of rotating gravastars with a physical surface localized on the BH horizon, whose results it is hoped 
will suggest approaches to a solution of the rotating gravastar interior, beyond the perturbative framework utilized in this paper. This requires a better understanding of the equation of state appropriate for rotating gravitational condensates. 

No attempt has been made in this paper to present either a microscopic theory of the surface stresses, which requires 
the incorporation of quantum vacuum polarization effects such as those discussed in \cite{MottolaVaulin:2006,EMZak2010}, 
nor a detailed model of the dynamical formation of rotating gravastars with $p=-\r$ cores from positive pressure matter. 
See for example~\cite{Beltracchi:2018ait} for preliminary study of a model of gravastar formation. These important
issues remain open to further investigation.

\vspace{3mm}
\centerline{\bf Acknowledgement}
\vspace{3mm}

The authors thank C. Posada for private communication confirming our understanding of the relation between his work and the present one,
as discussed in Sec.~\ref{Sec:SumDisc}. P.G. was partially supported by NSF grant No. PHY-2014075, and is very grateful to Prof.\ Masahide Yamaguchi for his generous support under JSPS Grant-in-Aid for Scientific Research Number JP18K18764 at the Tokyo Institute of Technology. E. M. acknowledges support from LANL LDRD grant No. 20200661ER.

\vspace{-3mm}
\bibliographystyle{apsrev4-2}
\bibliography{posadaeqs}

\appendix

\section{Einstein Tensor and Eqs.~to Second Order in the Angular Momentum}
\label{Sec:EinTens}

With the form of the line element (\ref{kingmet}) the nonvanishing components of the Einstein tensor are:
\begin{align}
&G^t_{~t}=-3H^2+ \Bigg[-\frac{2 m_0'}{r^2} + \frac{ j^2 r^2\!}{6}\, \w^{\prime 2} + \frac{ j^2 r\!}{3}\,\w\left(r \w''+4 \w'\right)\Bigg]\nn
&+ 2 P_2(\cos\th)\Bigg[h   k_2'' +\frac{\left(rh'+ 6h\right)\!}{2r}\,  k_2' -\frac{m_2'}{r^2} -\frac{2 k_2}{r^2} 
- \frac{3 m_2}{r^3h}-\frac{j^2 r^2}{12}\,\w^{\prime 2} -\frac{j^2 r\!}{6}\,\w  \left(r \w''+4\w'\right)\!\Bigg] 
\label{Gtt}\\
&G^r_{~r}=-3H^2+ \Bigg[\frac{2 h\!}{r}\, h_0' -\frac{2\left(r h'+h\right)\!}{h r^3}\,m_0+\frac{1}{6} j^2 r^2\w^{\prime 2}\Bigg]\nn
&+ 2 P_2(\cos\th)\Bigg[\frac{(rh' +2h)\!}{2r}\, k_2' + \frac{h}{r}\,h_2'  - \frac{3 h_2}{r^2} -\frac{2 k_2}{r^2}-\frac{\left(rh' + h\right)\!}{r^3h}\, m_2
-\frac{j^2 r^2\!}{12}\, \w^{\prime 2}\Bigg]
\label{Grr}\\
& \frac{G^r_{~\th}}{r^2h} =G^\th_{~r}= \frac{3 \sin\th\cos\th}{r^2}\left[ h_2'+k_2' +\frac{\!(rh'-2h)\!}{2rh }\,h_2 - \frac{\!(rh' + 2h)\!}{2r^2h^2 }\, m_2 \right]
\label{Gthr}\\
&G^\th_{~\th}= -3H^2+ \Bigg[h h_0'' + \frac{(3r h' +2h)\!}{2r}\, h_0' - \frac{(rh' + 2h)\!}{2r^2h}\,m_0' - \frac{h_2}{r^2} - \frac{m_2}{r^3h} \nn
&\hspace{4cm}+  \left( \frac{r^2h^{\prime 2}\!}{h} - 2r^2h'' - rh' +2h \right) \frac{m_0}{2r^3h} - \frac{j^2r^2}{6} \, \w^{\prime 2}\Bigg]\nn
& + P_2(\cos \th) \Bigg[h(h_2'' + k_2'') +\frac{(3rh' + 2h)\!}{2r} \,h_2' + \frac{(rh' + 2h)}{r} k_2' - \frac{(rh' + 2h)} {2r^2h} m_2'  - \frac{2h_2}{r^2} \nn
& \hspace{3cm}  + \left( \frac{ r^2h^{\prime 2}}{h} - 2 r^2 h'' -r h' + 2h -4\right) \frac{m_2}{2r^3h} + \frac{j^2 r^2}{6} \w^{\prime 2}\Bigg]
\label{Gthth}\\
&G^\f_{~\f}=-3H^2+  \Bigg[h h_0'' + \frac{(3r h' +2h)\!}{2r}\, h_0' - \frac{(rh' + 2h)\!}{2r^2h}\,m_0' + \frac{h_2}{r^2} + \frac{m_2}{r^3h} \nn
&\hspace{2cm}+  \left( \frac{r^2h^{\prime 2}\!}{h} - 2r^2h'' - rh' +2h \right) \frac{m_0}{2r^3h} - \frac{j^2r^2}{2} \, \w^{\prime 2}  -\frac{j^2 r\!}{3}\,\w  \left(r \w''+4\w'\right)\!\Bigg]\nn
    &+ P_2(\cos \th) \Bigg[h(h_2'' + k_2'') +\frac{(3rh' + 2h)\!}{2r} \,h_2' + \frac{(rh' + 2h)}{r} k_2' - \frac{(rh' + 2h)} {2r^2h} m_2'  - \frac{4h_2}{r^2} \nn
& \hspace{1cm}  + \left( \frac{ r^2h^{\prime 2}}{h} - 2 r^2 h'' -r h' + 2h -8\right) \frac{m_2}{2r^3h} + \frac{j^2 r^2}{2}  \w^{\prime 2}+\frac{j^2 r\!}{3}\,\w  \left(r \w''+4\w'\right)\!\Bigg]
\label{Gff}
\end{align}
where the terms in the large square brackets are second order in the angular momentum $J/M^2$, and
\begin{align}
G^t_{~\f}&=-\frac{j^2 r   \sin ^2\!\th}{2}  \left(r \w''+4 \w'\right)\\
G^\f_{~t}&=\frac{ h}{2 r}\left(r \w''+4\w'\right)
\label{Gft}
\end{align}
are first order in this parameter and somewhat simpler. The difference
\begin{align}
G^\th_{~\,\th} -  G^\f_{~\,\f}= -\frac{2\,}{r^2\!} \left[h_2 + \frac{m_2}{rh}  - \frac{j^2r^4}{6} \, \w^{\prime 2}  -\frac{j^2 r^3\!}{6}\,\w  \left(r \w''+4\w'\right) \right]\Big(1- P_2(\cos\th)\Big)
\label{Gdiff}
\end{align}
is particularly simple as well. In these expressions the prime $^\prime \equiv d/dr$ and we have used the zeroth order nonrotating gravastar solution (\ref{sphstat2}). 

The Einstein equations $G^\m_{~\,\n} = 8 \pi T^\m_{~\,\n} = -8\pi \EE\, \d^\m_{~\,\n}$ for the stress energy tensor (\ref{EMT}) are easily obtained from (\ref{Gtt})-(\ref{Gdiff}). 
Since $\na_\m G^\m_{~\,\n} \!=\!0$,\, $\EE= \r + \EE_0$ must be a constant. The terms in (\ref{Gtt}) and (\ref{Grr}) with no $\cos\th$ dependence give the monopole equations~\eqref{masterm0} and~\eqref{masterh0}, the latter in the form
\begin{align}
& \big(r-2m\big)\frac{dh_0}{dr} = \frac{( 1-2m')\! }{(r-2m)} \ m_0 - 4\pi r^2 \EE_0 -  \frac{j^2r^4}{12} \bigg( \frac{d\w}{dr} \bigg)^{\!2} 
\end{align}
after using $rh = r - 2 m$ from (\ref{sphstat2}), and with $m' \equiv dm/dr$. These equations may be compared to Eqs.~(18) and (19) of Ref.~\cite{Chandrasekhar:1974}, taking 
account of the constant values of $j$, $p=-\r$, and $\EE_0$ in the energy-momentum tensor (\ref{EMT}), as well as $\varpi' =\w'$. The vanishing of the difference $G^\th_{~\,\th} \!- G^\f_{~\,\f}$ in (\ref{Gdiff}), of the off-diagonal components $G^\th_{\ \,r}$ in (\ref{Gft}), and of the $P_2(\cos \th)$ terms of $G^r_{\ \,r}$ in (\ref{Grr}) give
\begin{subequations}
\begin{align}
&\hspace{3.6cm}h_2 + \frac{m_2}{rh}  - \frac{j^2r^4}{6} \, \w^{\prime 2} = 0 \label{masterdiff}\\
&\hspace{1.5cm}h_2'+k_2' +\frac{\!(rh'-2h)\!}{2rh }\,h_2 - \frac{\!(rh' + 2h)\!}{2r^2h^2 }\, m_2 = 0\label{masterGthr} \\
&(rh' +2h)\, k_2' + 2h \,h_2'  - \frac{6 h_2}{r} -\frac{4 k_2}{r}-\frac{2\left(rh' + h\right)\!}{r^2h}\, m_2 -\frac{1}{6}\,j^2 r^3 \w^{\prime 2} = 0 \label{masterGrr}
\end{align}
\label{masterquad}\end{subequations}
since all of these must vanish by Einstein's equations for the stress tensor (\ref{EMT}) with $\EE_2=0$. Solving (\ref{masterdiff}) for $m_2$ gives~\eqref{m2xx}, which when substituted into (\ref{masterGthr})-(\ref{masterGrr}) allows these to be written
\begin{subequations}
\begin{align}
r\big(r-2m\big) \frac{d}{dr} \big(h_2 + k_2\big) + 2\, \big(m-rm'\big) \, h_2 &= \sdfrac{1}{6}\big(r -m -rm' \big)\, j^2 r^4\w^{\prime 2}\\
r\big(r-2m\big) \frac{d}{dr} \big(h_2 + k_2\big)  + r \big(m-rm'\big)  \frac{d k_2}{dr}  -2r\big(h_2 &+ k_2\big)  -2 r m' \,h_2  =  \sdfrac{1}{12} \big(3-4m'\big)\, j^2 r^5\w^{\prime 2} 
\end{align}
\label{quadhk}\end{subequations}
after rearranging. Introducing the notation $v_2=h_2+k_2$, eliminating $k_2$ in favor of $v_2 -h_2$, and using $(m-rm' )= r(r-2m)\, \n_0'$,
the two coupled first order Eqs.~(\ref{quadhk}) may also be expressed as Eqs.~\eqref{v2xx} and~\eqref{h2xx}, which are directly comparable to Eqs.~(25) and (26) in~\cite{Chandrasekhar:1974}.
The remaining Einstein 
equations involving the second derivative terms in (\ref{Gtt})-(\ref{Gff}) are then satisfied automatically by the Bianchi identities.

\section{Weyl sector of the interior solution}
\label{Sec:Weyl}

The curvature tensors of our exterior solution are well known, since it is the Hartle-Thorne expansion of the Kerr metric. The curvature tensors of our interior solution 
are less well known, and we collect some information about them in this Appendix.

The Ricci sector of our interior solution is de Sitter like (albeit with a possible second order constant density shift $\EE_0$), but the full Riemann tensor has differences 
of first and second order in the perturbations. Of the integration constants, the constants $W_2,A_I,\d M_I$, and $B_I$ show up only in the Weyl sector, $\EE_0$  shows 
up only in the Ricci sector, and $W_1$ and $C$ do not show up in either sector and hence do not contribute directly to the curvature of the interior solution. 

A convenient method of examining the Weyl tensor is to use of the Petrov $Q$ matrix~\cite{Petrov:1954,Stephani:2003tm}. We derive the $Q$ matrix components 
for an arbitrary stationary axisymmetric metric of the form \eqref{axisymstat} in Appendix B of \cite{axihorizon}. The full $Q$ matrix components for our interior, and 
hence the Weyl tensor components,  are functions of $r,\theta$ and of the integration constants $W_2,A_I,\d M_I$, and $B_I$. Setting $B_I=0$ as in the finiteness condition~\eqref{BIfix}, the nonzero independent components of the $Q$ matrix are:
\vspace{1mm}
\begin{subequations}
\label{Qmat}
 \begin{align}
     Q_{11}=&\frac{6 i W_2 \cos \th}{r^4}+\frac{2\left( r^3  \d M_I- W_2^2\right)}{ r^6}\nonumber\\&-\frac{P_2(\cos\theta)}{ r^6 \RH^2} \Bigg[4 
     A_Ir \RH^2 \left(r^2-3 \RH^2\right)+2 W_2^2
   \left(15 r^2-11 \RH^2\right)\Bigg]
   \\
   Q_{21}=&\frac{3 i W_2 \sin \th \sqrt{\RH^2-r^2}}{r^4 \RH}+\frac{3\sqrt{\RH^2-r^2} \sin \th \cos \th}{ r^6 \RH^3
   }\Bigg[ 4 
   A_Ir \RH^4-5 W_2^2
   \left(r^2-2 \RH^2\right)\Bigg]
   \\
   Q_{22}=&-\frac{3 i W_2 \cos \th}{r^4}+\frac{1}{ r^6}\Bigg[ 
  A_Ir \left(\RH^2-r^2\right)- r^3 \d M_I +W_2^2
   \left(6-\frac{5 r^2}{\RH^2}\right)\Bigg]+\nonumber\\&\frac{P_2(\cos\theta)}{ r^6 \RH^2 }\Bigg[4 W_2^2 \left(5r^2-4\RH^2\right)+ 
   A_I r \RH^2 \left(3r^2-7\RH^2\right)\Bigg]
   \\
   Q_{33}=&-\frac{3 i W_2  \cos \th}{r^4}+\frac{1}{r^6}\Bigg[
  A_I r \left(r^2-\RH^2\right)-r^3 \d M_I+\frac{5 r^2 W_2^2}{\RH^2}-4
   W_2^2\Bigg]+\nonumber\\&\frac{P_2(\cos\theta)}{ r^6 \RH^2}\Bigg[ 
   A_Ir \RH^2 \left(r^2-5\RH^2\right)+2 W_2^2 \left(5r^2-3\RH^2\right) \Bigg]
 \end{align}
 \end{subequations}
 To lowest order in the rotation, the $Q$ matrix vanishes, in accordance with the fact that de Sitter space has a vanishing Weyl tensor. 
Note that $W_2$ causes a first order correction in the ``magnetic" part of the Weyl tensor (imaginary part of the $Q$ matrix). 
The correction to the ``electric" part (real part of the $Q$ matrix) is of second order. The $Q$ matrix contains divergences at $r=0$ 
for nonzero $J$, and it is impossible to choose $W_2$ and $\d M_I$ to cancel all diverging terms in the Weyl tensor at the origin.
Since the Ricci tensor and scalar as determined by the stress-energy (\ref{EMT}) is finite, the singular behavior of the Weyl tensor
at the origin is the same as that of the corresponding components of the Riemann tensor. If one eliminates all the $\d$-function 
contributions at the origin to the Komar mass and angular momentum by setting $W_2=\d M_I= 0$ as in (\ref{W2E0}), and uses~\eqref{AIfix} for $A_I$, some of the divergences 
at $r=0$ are removed and (\ref{Qmat}) simplifies considerably to
\begin{subequations}
\label{Qmat0}
 \begin{align}
Q_{11}&=\frac{8J^2 P_2(\cos\th)(r^2-3\RH^2)}{r^5\RH^3}\\
Q_{21}&=-\frac{24 J^2}{ r^5\RH^2} \sqrt{\RH^2-r^2} \, \sin\th \cos\th\\
Q_{22}&=\frac{2J^2(r^2-\RH^2)}{r^5\RH^3}+\frac{2J^2 P_2(\cos\th)(7\RH^2-3r^2)}{r^5\RH^3}\\
Q_{33}&=\frac{2J^2(\RH^2-r^2)}{r^5\RH^3} +\frac{2J^2P_2(\cos\th)(5\RH^2-r^2)}{r^5\RH^3}.
\end{align}
\end{subequations}
\vspace{-1cm}

\section{Null coordinates and conformal diagram}
\label{Sec:Conform}
 
 In the spherically symmetric case, one can move to Eddington-Finkelstein coordinates $v$ or $u$ using
\begin{align}
    dv=dt+\frac{1}{\sqrt{f h}} dr,\qquad\text{or}\qquad
    du=dt-\frac{1}{\sqrt{f h}} dr,
\end{align}
respectively.
Note that in Eddington-Finkelstein coordinates the metric
\begin{align}
    ds^2=-f du^2-2 \sqrt{\frac{f}{h}}\,du\, dr+r^2 \big(d\th^2 + \sin^2\!\th \, d\f^2\big)
\end{align}
has a discontinuity on the null $r=\RH$ hypersurface where $f(\RH)=0$.

\begin{figure}[h]
    \centering
    \includegraphics[width=10cm]{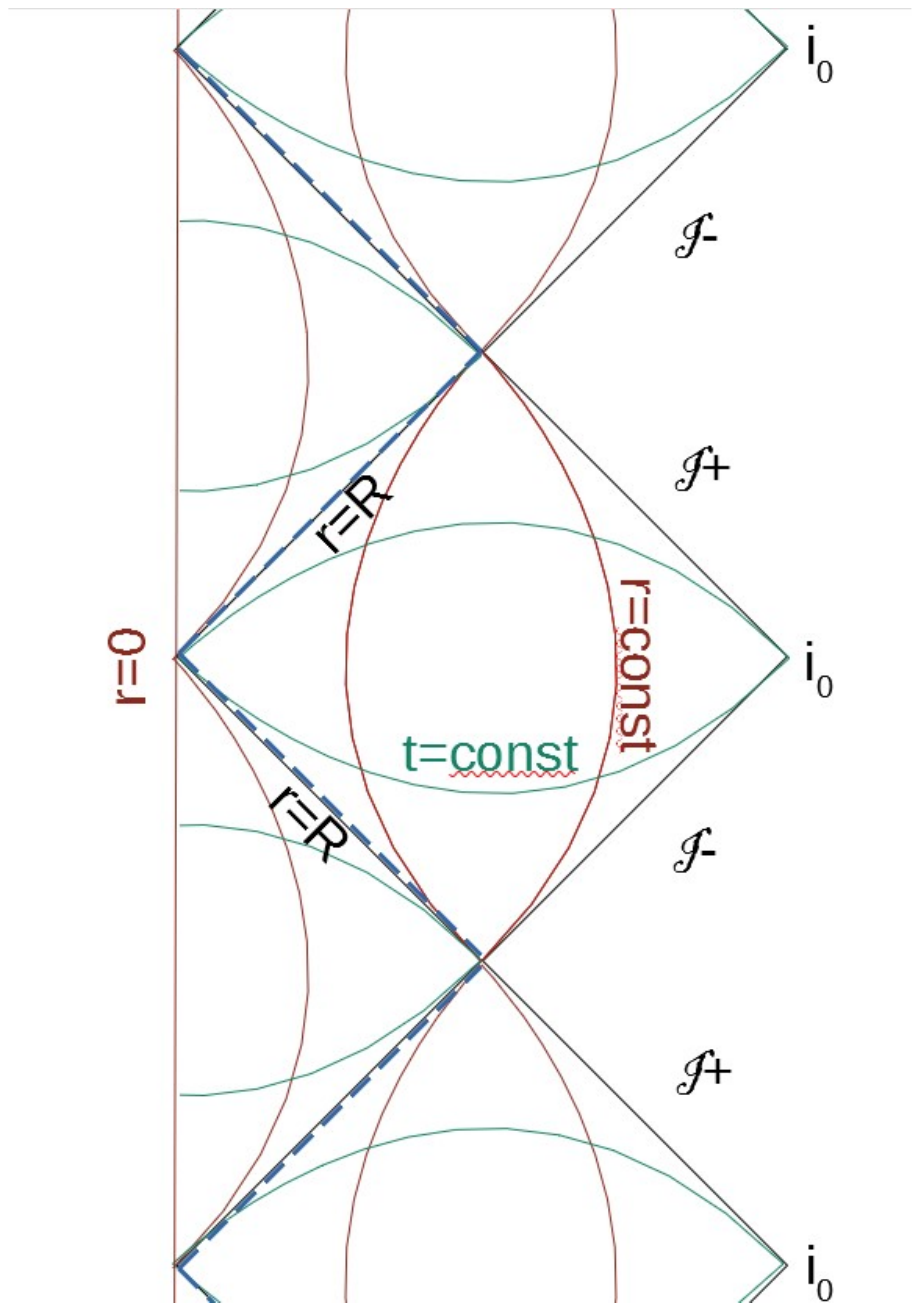}
    \caption{Conformal diagram for the static gravastar. Schwarzschild exterior (``squares'') and de Sitter static patch interior (``triangles'') diagrams are joined at the $r=\RH$ null hypersurface and may be tiled in a structure similar to the extremal Reissner-Nordstrom black hole. Unlike the Reissner-Nordstrom spacetime, the static gravastar is regular at $r=0$ but has a singularity associated with the surface at $r=\RH$ (shown in the diagram by a dashed blue line). Sketches of constant $r$ hypersurfaces and of constant $t$ hypersurfaces are shown.  }
    \label{conformal}
\end{figure}

Because the interior is a de Sitter space and the exterior is Schwarzschild, we may construct a conformal diagram, which we sketch in Figure \ref{conformal}. It is reminiscent of the conformal diagram for the extremal Reissner-Nordstrom solution, although now $r=0$ is regular, but the null surface is associated with a Dirac $\delta$-function energy-momentum distribution and hence is singular.

\end{document}